\documentclass[fleqn,usenatbib]{mnras}

\usepackage{newtxtext,newtxmath}

\usepackage[T1]{fontenc}

\DeclareRobustCommand{\VAN}[3]{#2}
\let\VANthebibliography\thebibliography
\def\thebibliography{\DeclareRobustCommand{\VAN}[3]{##3}\VANthebibliography}

\usepackage{graphicx}	
\usepackage{amsmath}	
\usepackage{bm}
\usepackage{subcaption}
\usepackage{multirow}
\usepackage{fix-cm}

\usepackage{scalerel,tikz}
\usetikzlibrary{svg.path}
\definecolor{orcidlogocol}{HTML}{A6CE39}
\tikzset{orcidlogo/.pic={
 \fill[orcidlogocol] svg{M256,128c0,70.7-57.3,128-128,128C57.3,256,0,198.7,0,128C0,57.3,57.3,0,128,0C198.7,0,256,57.3,256,128z};
 \fill[white] svg{M86.3,186.2H70.9V79.1h15.4v48.4V186.2z}
 svg{M108.9,79.1h41.6c39.6,0,57,28.3,57,53.6c0,27.5-21.5,53.6-56.8,53.6h-41.8V79.1z M124.3,172.4h24.5c34.9,0,42.9-26.5,42.9-39.7c0-21.5-13.7-39.7-43.7-39.7h-23.7V172.4z}
 svg{M88.7,56.8c0,5.5-4.5,10.1-10.1,10.1c-5.6,0-10.1-4.6-10.1-10.1c0-5.6,4.5-10.1,10.1-10.1C84.2,46.7,88.7,51.3,88.7,56.8z};
}}
\newcommand\orcidicon[1]{\href{https://orcid.org/#1}{\mbox{\scalerel*{
\begin{tikzpicture}[yscale=-1,transform shape]
\pic{orcidlogo};
\end{tikzpicture}
}{|}}}}

\title[Blind source separation of the stellar halo]{Blind source separation of the stellar halo}

\author[E. Y. Davies et al.]{Elliot Y. Davies~\orcidicon{0000-0001-5996-4072}$^{1}$\thanks{E-mail: eyd20@cam.ac.uk}, Vasily Belokurov~\orcidicon{0000-0002-0038-9584}$^{1}$, Andrey Kravtsov~\orcidicon{0000-0003-4307-634X}$^{2}$, Stephanie Monty~\orcidicon{0000-0002-9225-5822}$^{1}$, \newauthor GyuChul Myeong~\orcidicon{0000-0002-5629-8876}$^{1}$, N. Wyn Evans~\orcidicon{0000-0002-5981-7360}$^{1}$ and Sarah G. Kane~\orcidicon{0000-0001-8411-1012}$^{1}$. \\
$^{1}$Institute of Astronomy, University of Cambridge, Madingley Road, Cambridge CB3 0HA, UK \\
$^{2}$Department of Astronomy and Astrophysics, The University of Chicago, Chicago, IL 60637, USA}

\date{Accepted XXX. Received YYY; in original form ZZZ}

\pubyear{2015}

\begin{document}
\label{firstpage}
\pagerange{\pageref{firstpage}--\pageref{lastpage}}
\maketitle

\begin{abstract}
The stellar halo of the Milky Way comprises an abundance of chemical signatures from accretion events and \textit{in-situ} evolution, that form an interweaving tapestry in kinematic space. To untangle this, we consider the mixtures of chemical information, in a given region of integral of motion space, as a variant of the blind source separation problem using non-negative matrix factorisation (NMF). Specifically, we examine the variation in [Fe/H], [Mg/Fe], and [Al/Fe] distributions of APOGEE DR17 stars across the $(E,L_z)$ plane of the halo. When 2 components are prescribed, the NMF algorithm splits stellar halo into low- and high-energy components in the $(E,L_z)$ plane which approximately correspond to the accreted and \textit{in-situ} halo respectively. We use these components to define a boundary between the \textit{in-situ} and the accreted stellar halo, and calculate their fractional contribution to the stellar halo as a function of energy, galactocentric spherical radius ($r$), height ($z$), and galactocentric cylindrical radius ($R$). Using a stellar halo defined by kinematic cuts, we derive a boundary in $(E,L_z)$ space where the halo transitions from \textit{in-situ} dominated to accretion dominated. Spatially, we find that this transition happens at $(r,z,R) \approx (8.7, 3.0, 8.1)$ kpc. We find that between 34\% to 53\% of the stellar halo's content is of accreted origin. Upon prescribing more components to the NMF model, we find evidence for overlapping chemical evolution sequences. We examine features within these components that resemble known substructures in the halo, such as \textit{Eos} and \textit{Aurora}. 
\end{abstract}




\section{Introduction}

Galactic stellar halos are a mess. An ambitious promise of Galactic Archaeology is to bring order to this chaos in the Milky Way. The trailblazing pre-{\it Gaia} papers \citep[see e.g.][]{HdZ2000} suggested that individual merger debris could be disentangled in the space of integrals of motion (IOM). If successful, this would allow us to verify our hierarchical structure formation theory \citep[][]{white1978core} and study long-gone, low-mass Galactic fragments. As the data and the simulations improved, this archaeological framework encountered several stumbling blocks and needed adjustments. For example, if several massive mergers contributed to the formation of the stellar halo, their IOM signals would likely overlap \citep[][]{jeanbabtiste2017kinematic} as the dwarfs sink and radialize in a dissipative two-body interaction with the host \citep[][]{Amorisco2017,vasiliev2022radialization}. In the Milky Way however, the data appears to favour a more quiescent accretion history dominated by a single significant ancient merger \citep[see e.g.][]{deason2013broken}, which can flood the local stellar halo with its debris thus overwhelming and suppressing signals from smaller sub-structures \citep[see discussion in][]{Deason2024}. 

The solution to the problem of the overcrowded IOM space came in the form of high-resolution, high-multiplex, wide-angle spectroscopy supplied by surveys such as the Apache Point Observatory Galactic Evolution Experiment \citep[APOGEE,][]{Majewski2017APOGEE} and Galactic Archaeology with HERMES \citep[GALAH,][]{Martell2017GALAH}. Precise elemental abundances available within these datasets can hypothetically be used as a chemical fingerprint to isolate and authenticate stars born in different environments, even if they occupy the same location in IOM space \citep[see][for the discussion of chemical tagging, although more in the context of the Galactic disc]{Freeman2002}. The first convincing example of applying detailed elemental abundances to chemically tag distinct stellar halo populations (in the Solar neighborhood) can be found in \citet{Nissen2010}. The two tight sequences in the $\alpha$-[Fe/H] space they uncovered intrigued many and inspired a plethora of related chemical studies \citep[e.g.][]{Hawkins2015,Bonaca2017, Fernandez-Alvar2018, Hayes2018}. It took the {\it Gaia} data \citep[][]{gaia2018gaia} to piece the puzzle together, connecting the low-$\alpha$ sequence to our Galaxy's last significant merger also known as {\it Gaia} Sausage/Enceladus \citep[GS/E, e.g.,][]{belokurov2018coformation,Haywood2018,helmi2018merger}, and the high-$\alpha$ sequence to stars in the old, splashed-up MW disc \citep[e.g.,][]{Bonaca2017, Gallart2019, Dimatteo2019,belokurov2020biggest}. 

For the task of unpicking the jumble of the stellar halo, there is undeniable power in combining chemical and orbital information. In fact, on their own, neither orbits nor abundances are likely to succeed. Most recently, this was demonstrated in a number of experiments conducted to cleanly isolate the stars belonging to the GS/E \citep[e.g.][]{mackereth2019origin,Das2020,Feuillet2021,Buder2022,Carrillo2024}. Typically, simple, smooth decision boundaries are drawn in selected orbital and chemical dimensions. However, this approach is clearly unsatisfactory, as no projection of the parameter space has yet been identified where the GS/E stars are completely separable from the rest of the halo. To circumvent the hurdle of overlapping populations, Gaussian Mixture Modelling has been employed to describe the distribution of stars in the space of chemical abundances \citep[see e.g.][]{Das2020,Buder2022,myeong2022milky} and IOMs \citep[][]{myeong2022milky,callingham2022chemo}. GMM gracefully handles intersecting data clouds, but unfortunately, this strategy is not physically motivated as chemical distribution functions are rarely Gaussian. Instead, in elemental abundance spaces, stellar populations from the same progenitor form long, narrow sequences with asymmetric density distributions shaped by the physics of nucleosynthesis as dictated by models of galactic chemical evolution \citep[][]{Spitoni2017,Andrews2017,Weinberg2017,Kobayashi2020}. One example of how GMM inherently struggles to handle high-quality abundance measurements of stars spanning a large range of metallicity can be found in \citet{myeong2022milky}. They find that the uninterrupted GS/E chemical sequence requires two distinct GMM components for accurate representation. This is due to a small but noticeable deflection in the GS/E sequence, as seen in most abundance ratios relative to iron as a function of [Fe/H], caused by the increased contribution of Type Ia supernovae.

Despite appearing as a tangled mess, the local stellar halo exhibits a clear trend: elemental abundances show a pronounced evolution with orbital energy. This is first pointed out in \citet{myeong2018sausage} who notice that Galactic globular clusters (GCs) separate into two distinct groups at a "critical energy". \citet{belokurov2022dawn} and \citet{Belokurov2023nitrogen} focus instead on field stars in APOGEE and using aluminum abundance as a chemical fingerprint -- inspired by the pioneering work of \citet{Hawkins2015} -- arrive at a similar conclusion. They show that a sharp boundary can be drawn in the space of energy $E$ and the vertical component of angular momentum $L_z$. Higher-energy (less bound) stars have low levels of [Al/Fe], while lower-energy stars, on average, have higher [Al/Fe]. 

If aluminium abundance is interpreted as a tracer of early chemical evolution, then the low-energy stars must have formed in an environment with a rapid enrichment, typical of a massive progenitor, while the high-energy stars come from a system with a slower pace of star formation. Further support for this hypothesis can be found in the analysis of \citet{Hassselquist2021} who show that all MW dwarf satellites with chemistry available in APOGEE display low levels of [Al/Fe]. Most recently, \citet{Monty2024} reveal that the global orbital trends in the halo are not limited to [Al/Fe]. They show that $\alpha$-element abundances, for example [Si/Fe], and neutron-capture $r$-process elements, such as [Eu/Fe], as well as their ratio [Eu/$\alpha$] also show low/high energy separation. However,  \cite{Monty2024} highlight that such a correlation between [Eu/$\alpha$] and orbital energy is subject to a metallicity selection. This is because both $\alpha$ and $r$-process abundances evolve with metallicity, thus assuming that the above chemo-dynamical pattern persists at all metallicities is an over-simplification. The same must be true for [Al/Fe].

From a dynamical perspective this chemical behaviour makes sense, since the stellar halo can be broadly divided into \textit{in-situ} and accreted stars. In-situ stars tend to be more gravitationally bound due to early gas cooling, while accreted stars arrive with high orbital energy that is gradually dissipated through dynamical friction. This process, particularly in high-mass-ratio mergers, also leads to significant angular momentum loss and radialization of accreted stars \citep[][]{Amorisco2017,vasiliev2022radialization}. Simulations confirm that such radialization is common in the formation of the stellar halo, particularly in GS/E-like mergers \citep[][]{Villalobos2008,belokurov2018coformation,naidu2021reconstructing,belokurov2023energy}. Among tailored GS/E simulations, \citet{amarante2022gastro} stands out. Their models, which incorporate gas dynamics and star formation, suggest that the mass of the GS/E progenitor strongly influences the energy distribution of its debris. Namely, lower-mass dwarfs remain at higher orbital energies, aligning well with observational constraints \citep[e.g.][]{Lane2023}. The lightest dwarf's unbound stars produce an eccentricity distribution with the most pronounced pile-up close to $e=1$, noticeably sharper than the broad distributions of the other dwarfs' debris. The emerging picture suggests that GS/E was a moderately massive merger, leaving behind debris at relatively high orbital energies. Since then, no similarly massive accretion events have occurred (apart from the ongoing Sgr and LMC interactions), implying that the bulk of the low-energy stellar halo formed differently—either through an earlier merger \citep[e.g.][]{kruijssen2020kraken,Horta2021} or from the in-situ pre-disk population Aurora \citep[][]{belokurov2022dawn}.

Therefore, a picture emerges in which the main contributor to the local stellar halo was of modest stellar mass and as a result, its debris is currently levitating at relatively high levels of orbital energy. With no strong evidence for any subsequent massive accretion events since the GS/E, outside of the Sgr and LMC that are still ongoing \citep[but see][for a different interpretation]{donlon2022local}, the bulk of the low-energy stellar halo must have formed differently and/or at earlier epochs. Indeed, several recent studies advocate for a substantial pre-GS/E merger whose debris is located at sub-Solar energy levels, thus populating the very heart of the stellar halo. More specifically, \citet{kruijssen2020kraken} and \citet{Forbes2020} identify a group of low-energy Globular Clusters which they associate with a massive and ancient event dubbed Kraken or Koala \citep[see also][]{massari2019origin}. \citet{Horta2021} rely on detailed chemistry from APOGEE, namely [Mg/Mn] and [Al/Fe], to argue for equally old and equally immense merger they call Heracles. Whether the same event or not, Kraken/Koala/Heracles accretion remnant is argued by these authors to be the dominant component of the Galactic inner stellar halo.

An alternative hypothesis is proposed by \citet{belokurov2022dawn} who argue that the bulk of the low-energy halo is composed of stars from Aurora, the pre-disk in-situ MW population. Their main evidence for Aurora is chemical: it is revealed by the global correlation between [Al/Fe] and orbital energy discussed above and characterised by elevated scatter in many other elemental abundances. In a follow-up study, \citet{Belokurov2023nitrogen} argue that some or most of this abundance scatter is due to a large contribution of globular cluster stars during the pre-disk, Aurora phases of MW evolution, in which up to $\approx 50\%$ of starformation occurred in massive bound star clusters. 

Finally, \citet{belokurov2024insitu} use the IOM boundary derived in \citet{Belokurov2023nitrogen} using the global [Al/Fe] pattern and classify all MW GCs as either in-situ or accreted. They show that as much as $\sim$2/3 of all currently observed Galactic GCs were formed in situ. Furthermore, they argue that the bulk of the GCs at low orbital energies are likely of in-situ origin, but point out that a small number of accreted objects among them cannot be ruled out. Recent work by \citet[][]{chen2024galaxy} similarly found that 60$\%$ of GCs in the MW can be categorised as \textit{in-situ}, using clustering methods calibrated on galaxy formation simulations of MW-sized galaxies.   These findings are supported by models of galaxy formation: in the FIRE-2 numerical simulations \citep[see][]{Wetzel2023}, the inner stellar halo is dominated by an in-situ pre-disk component similar to Aurora \citep[see][]{belokurov2022dawn} and the orbital energy distributions of low-metallicity stars show strong in-situ/accreted bi-modality comparable to that observed in the Galaxy \citep[][]{Belokurov2023nitrogen,belokurov2024insitu}. 

It is sometimes argued that the distinction between in-situ and accreted stars becomes increasingly blurred, or may even disappear entirely, as we examine the earliest observable epochs in the Galaxy's history. For example, \citet{Horta2024} study the outputs of the FIRE-2 simulations and find that in $\sim40\%$ of the hosts, the proto-Galaxy is composed of at least two "building blocks" with similar masses and similar chemical and orbital properties. While in-depth studies of the earliest stages in the MW's formation are limited to a relatively small number of accretion histories probed in simulation suites like IllustrisTNG \citep[][]{nelson2019illustris, pillepich2019first}, Auriga \citep[][]{grand2017auriga}, FIRE-2 \citep[][]{wetzel2017reconciling} and ARTEMIS \citep[][]{font2020artemis}, a clear pattern starts to emerge. 

\citet{belokurov2022dawn} argue that a new tight constraint on the assembly of the MW halo can be obtained from the measurements of the so-called "spin-up" transition which occurs when the mode of the stellar velocity distribution shifts quickly up, signifying the emergence of the stable stellar disk \citep[see also][]{conroy2022birth,Xiang2022}. In simulations only $\sim10\%$ of MW-mass hosts experience spin-up transitions at metallicity as low as estimated in the Galaxy \citep[see][]{belokurov2022dawn,Semenov2023,Dillamore2023spin,chandra2024three, zhang2024existence}. Crucially, these MW analogs assemble their DM mass the earliest compared to other systems of comparable halo mass at $z=0$. Moreover, to keep the ancient stellar disk intact, these hosts undergo noticeably quieter latter stages of their evolution  \citep[][]{Dillamore2023spin}.

Given the growing evidence for an increasingly messy Galactic stellar halo, we must investigate the MW's content with tools that do not enforce unnecessary priors. Moreover, we believe it is essential to make use of all the machine-learning tools that we have available to use. In this paper, we address the question of identifying the major components of the local stellar halo while making as few assumptions as possible. Specifically, we treat the problem of finding sub-population in the Galactic halo a variant of the so-called blind source separation (BSS). 

In essence, BSS involves separating a population of source signals from a population of mixed signals, without much (or any) knowledge of how the sources were mixed, or what they look like. The two assumptions of this approach are 1) the number of source signals and 2) the overall stellar distribution is a linear mixture of the assumed number of components. The linearity assumption is particularly justified given that we expect that the two components are shaped by independent physical processes, such as the evolution of {\it in-situ} and accreted stellar populations. Crucially, by treating the problem as a variant of BSS, different components can overlap and can have arbitrary shapes dictated by the data rather than defined {\it a priori} subjectively. 

While there are a range of machine learning methods that can be used to tackle BSS, we opt to use non-negative matrix factorisation (NMF) \citep[][]{Paatero1994,Antilla1995,Lee1999learning,Lee2001}. In brief, NMF aims to rewrite a single data matrix that consists of a population of mixed signals as a product of two matrices, $\boldsymbol{V} \approx \boldsymbol{WH}$, in order to break the mixed signals up into their un-mixed components. In essence, NMF decomposes presented data into a linear mixture of non-negative components. 

In this study, we apply NMF to a sample of APOGEE DR17 \citep[][]{APOGEEDR17} stars with full 6-dimension kinematic information. In many problems, including the problem of separating different physical components by analysing the density of stars in certain spaces of chemical abundances, non-negativity is a required assumption and ensures intuitive, easily interpretable results \citep[e.g.][]{gillis2014why}. This built in non-negativity makes NMF an attractive tool. Moreover, what makes NMF particularly appealing is that, unlike other methods, it does not require any other strong restrictive assumptions. Despite being developed for use in signal processing, NMF has already found utility in astrophysics \citep[e.g.][]{blanton2007kcorrections, allen2011strong,Ren2018}. 

By applying NMF to the APOGEE data we aim not only to come up with an unbiased break-down of the halo but also shed new light on its least understood components, such as the recently discovered Eos population \citep[][]{myeong2022milky,matsuno2024distinct}. Eos stars follow a highly unusual trajectory in the chemical abundance space, essentially connecting the cloud of the GS/E stars with the in-situ track. The exact nature of this population is currently unknown, but cosmological simulations have shown that an \textit{Eos}-like population can result after a major accretion event, as the merger causes a starburst amongst the inner \textit{in-situ} stars that are puffed up onto halo orbits \citep[][]{grand2020dual, renaud2021vinter}. With all this in mind, \textit{Eos} is theorised to be a population of stars born \textit{in-situ} from gas polluted by the GSE merger \citep[see also][]{An2023,Ciuca2024, chen2024dawn}, and is therefore a perfect example of the interconnected and complex nature of structures found within Milky Way's halo. 

This Paper is organised as follows. In Sec.~\ref{sec:data}, we present the data used in this study, detailing quality cuts and stellar halo cuts. In Sec.~\ref{sec:methods}, we discuss the methods use for analysing this data and explain NMF and BSS in more detail. In Sec.~\ref{sec:components} we present and discuss the resulting components from different NMF models, and in Sec.~\ref{sec:application} we apply these components to other datasets. In Sec.~\ref{sec:deepdive} we look into some of the specific unusual features within these components. Lastly, we summarise our findings in Sec.~\ref{sec:results}.

\section{Data}\label{sec:data}

\begin{figure*}
    \centering
    \includegraphics[width=0.9\textwidth]{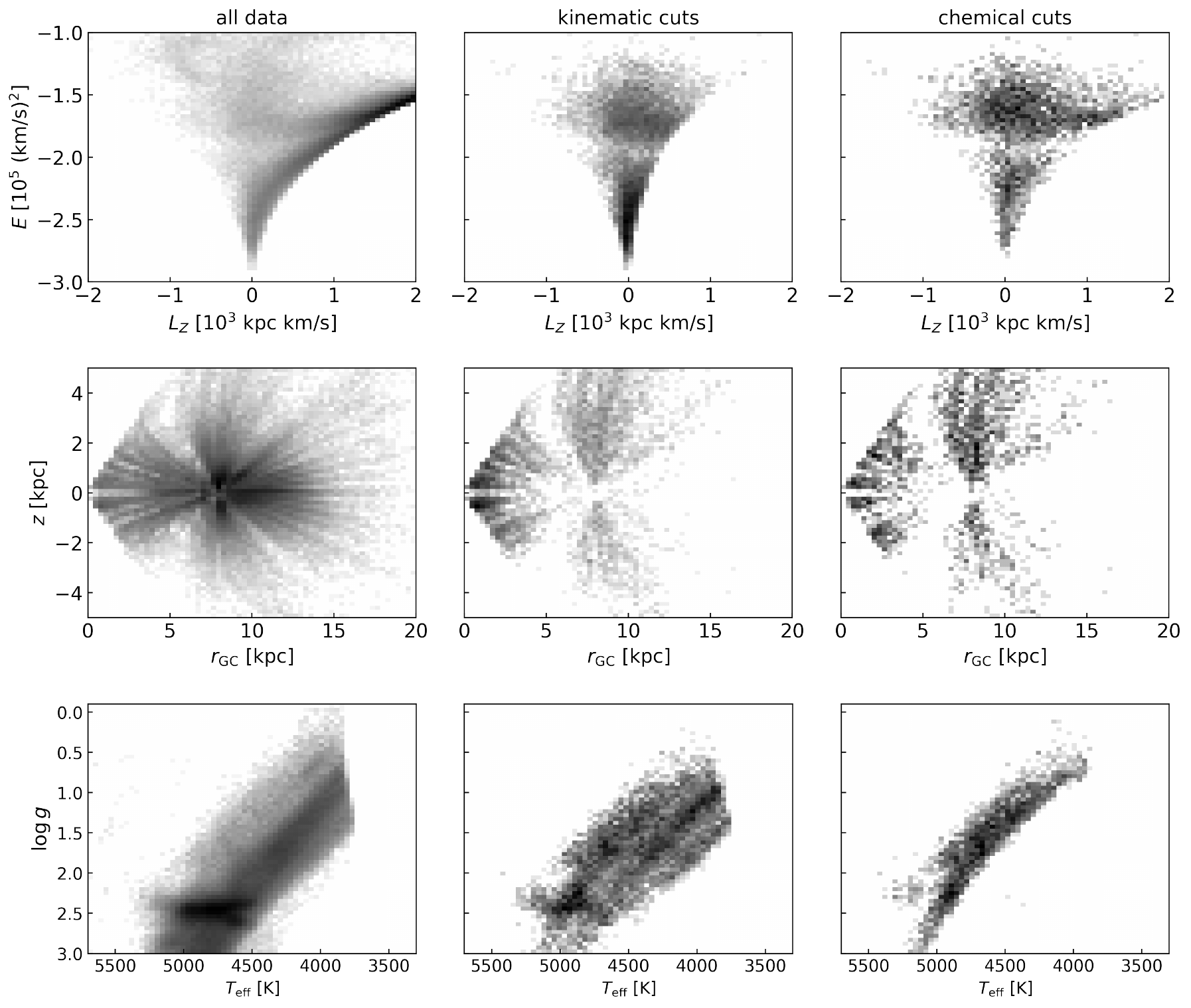}
    \caption{Dataset of APOGEE DR17 stars used for this study, shown in energy and angular momentum space (top row), spatial distribution (middle row), and as a Kiel diagram (bottom row). The leftmost panel shows all of the stars that passed our quality cuts. The central and rightmost panels illustrate our two selections of stars belonging to the stellar halo. The central panel shows a chemically pure stellar halo, whose stars are selected by a cut in eccentricity and angular momentum. The rightmost panel shows a kinematically pure stellar halo, whose stars are selected by a cut in metallicity. Specifically, our kinematic cuts require stars with $L_z >0$ have $e < 0.7$, and our chemical cuts require all stars have [Fe/H] < -1.1 dex.}
    \label{fig:fig1}
\end{figure*}

\begin{figure}
    \centering
    \includegraphics[width=0.9\columnwidth]{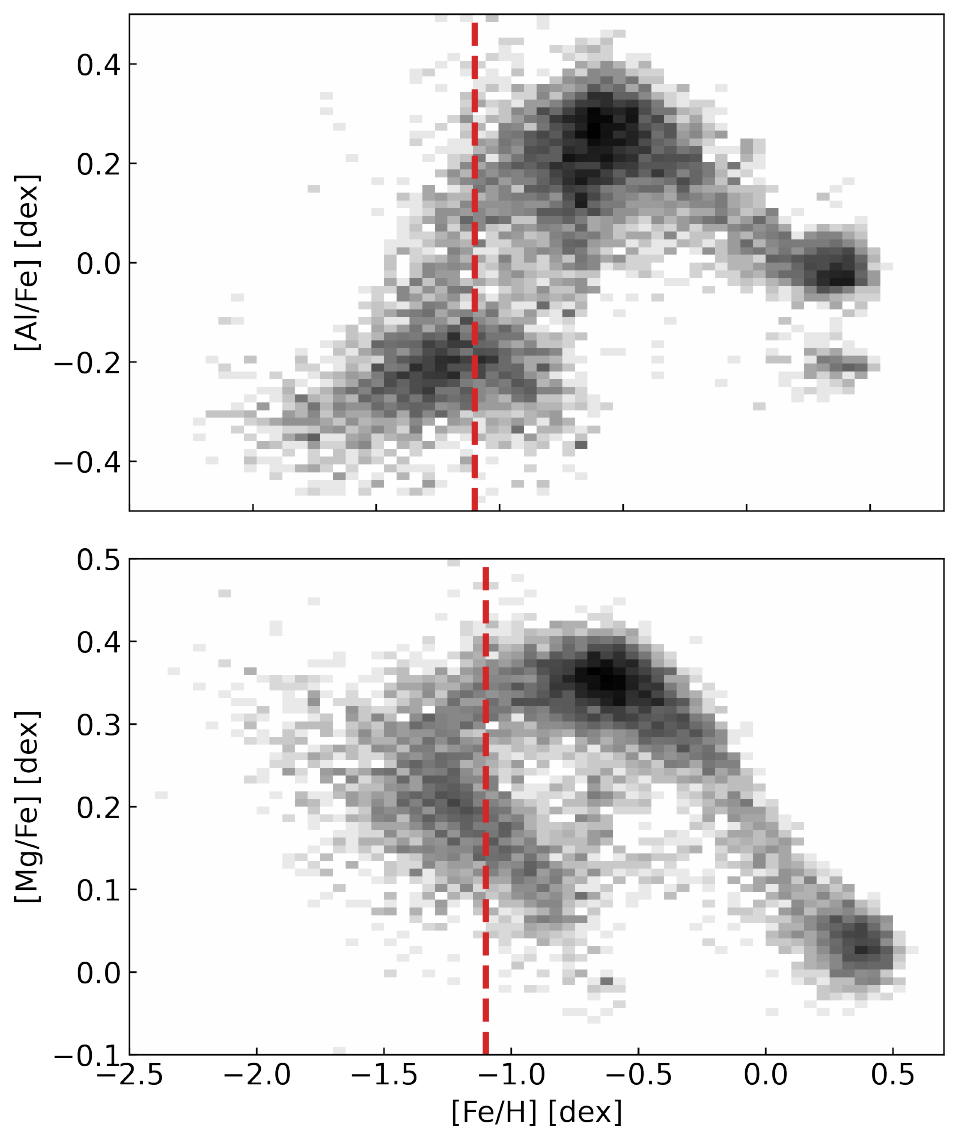}
    \caption{The resulting chemical spaces of APOGEE DR17 stars after the kinematic cuts have been applied. The red dashed line shows the chemical cut at [Fe/H] = -1.1 dex.}
    \label{fig:fig2}
\end{figure}

In this work we obtain all data from APOGEE DR17 \citep{APOGEEDR17}, specifically the AllStarLite catalogue supplemented with distances from the AstroNN Value-Added-Catalogue \citep[][]{leung2019deep}. Additionally, we utilise the catalogue of $\sim 1\times 10^6$ globular cluster (GC) member stars presented in \citet[][]{vasiliev2021gaia} to remove GCs from our dataset. All stars with a GC membership probability of 0.8 or greater are removed. Any orbital properties are calculated using \textsc{AGAMA} \citep[][]{vasiliev2019agama}, where we assume the Milky Way potential presented in \citet[][]{mcmillan2017mass} throughout. We also assume the default \textsc{Astropy} values for the solar position, and solar velocity, which are $r_{\odot} = 8.122$ kpc and $\boldsymbol{v}_{\odot} = (12.9, 245.6 7.78)$ km/s, respectively. This implies a solar z-angular momentum of $L_{z \odot} = 1995$ kpc km/s.

First, a series of quality cuts are made to clean the data. We remove all stars with \texttt{STAR\textunderscore BAD}, \texttt{TEFF\textunderscore BAD}, and/or \texttt{LOGG\textunderscore BAD} flag in \texttt{ASPCAPFLAGS}. Similarly we remove all duplicate and telluric data using the \texttt{EXTRATARG} bitmask. Additionally, we remove all stars associated with the Magellanic clouds by making use of the \texttt{magcloud} flag in \texttt{PROGRAMNAME}. For all elements \texttt{X}, we apply a cut of \texttt{X\textunderscore FE\textunderscore ERR}~$< 0.2$ and enforce \texttt{X\textunderscore FE\textunderscore FLAG}~$=0$. We also limit the sample to red giants by applying a \texttt{LOGG}~$<3.0$ cut. Moreover, we only select stars within heliocentric distance of 10 kpc from the Sun and enforce that the energy is of all particles is $E < 0$. 

Since this study is intended to be limited to the Galactic halo, we isolate the stellar halo by making further cuts to the data. To obtain the most agnostic results, we consider two different selections for the stellar halo. We create a ``chemically pure'' stellar halo by making cuts on kinematic space, and we create a ``kinematically pure'' stellar halo by making cuts on chemical space. Specifically, for the chemically pure stellar halo we keep stars either (a) with $L_z < 0$, or (b) with both $e > 0.7$ and $L_z>0$. In essence, we keep all retrograde stars, but retain only highly eccentric prograde stars. For the kinematically pure stellar halo, we cut out all stars with [Fe/H] > -1.1 dex. The above cuts leave us with a sample of $N=13147$ stars for the chemically pure halo and $N=5387$ for the kinematically pure halo. One major caveat of the these two samples is that they contain a large number of inner galaxy stars. Namely, both samples contain a large fraction of stars at galactocentric radii of less than 5 kpc. This is evident in Fig.~\ref{fig:fig1}, where we present the spatial distribution ($r_{\rm gc}, z$) of all samples used in the study. While these stars could belong to the halo, it is possible they belong to the central bulge. To not remove inner halo stars from our sample, we separately study the impact of the removal of these inner stars in Appendix~\ref{sec:appendix1}.

Given the proven utility of [Mg/Fe] and [Al/Fe] in separating the \textit{in-situ} and accreted components of the Galactic halo, we focus on these elements in our study, as well as [Fe/H]. The result of these cuts is presented in Fig.~\ref{fig:fig1} and Fig.~\ref{fig:fig2}. Evident across all three panels of Fig.~\ref{fig:fig1} is gap at $E \gtrsim -2.0 \times 10^5$ (km/s)$^2$, which is an artefact of the APOGEE selection function itself and not our cuts. The link between the inner galaxy and this artifact is addressed in Appendix~\ref{sec:appendix1}.

\section{Methods}\label{sec:methods}

In this section, we describe the background and detail of the NMF method, the processing of the data before it is input to the NMF, and some of the analysis methods that are applied to the NMF outputs.

\subsection{Blind source separation}

Blind source separation (BSS) is a method of untangling a mixture of signals into their component sources \citep[e.g., see][]{cardoso1998blind}. While most commonly thought of as a way to separate mixed audio signals \citep[also known as the cocktail party problem,][]{cherry1953some}, it is also widely used for the separation of images. It has already proved to be useful in astronomical applications \citep[e.g.,][]{blanton2007kcorrections, allen2011strong, Ren2018}. The method involves rewriting an $m \times n$ matrix containing the mixed data, $\boldsymbol{V}$, of the assumed number $p$ of mixed components into a product,
\begin{equation}\label{eq:nmf_factorisation}
    \boldsymbol{V} \approx \boldsymbol{WH},
\end{equation}
where $\boldsymbol{W}$ is an $m \times p$ matrix of weights, and $\boldsymbol{H}$ is an $p \times n$ matrix of reconstructed components.

\subsection{Non-negative matrix factorisation}

The algorithm that we use to conduct BSS in this work is a version of non-negative matrix factorisation (NMF), proposed by \citet[][]{Lee1999learning} and further developed in \citet{Lee2001}. NMF imposes a non-negativity constraint on $\boldsymbol{H}$ and $\boldsymbol{W}$, which is sensible for our application. Moreover, the number of components generated is less than the input ($p < m$) and so this method can be used as a dimension reduction tool. The aim of NMF is to minimise an appropriate cost function that measures the quality of the approximation in Eq.~\ref{eq:nmf_factorisation}, with respect to $\boldsymbol{W}$ and $\boldsymbol{V}$ such that $\boldsymbol{W},\boldsymbol{V} \geq 0$. Such an appropriate cost function is the euclidean distance:
\begin{equation}\label{eq:euclid_distance}
    ||\boldsymbol{V} - \boldsymbol{WH}|| = \sqrt{\sum_{ij}\left(V_{ij} - \sum_{k}W_{ik}H_{kj}\right)^2}
\end{equation}
It is common to begin by initialising $\boldsymbol{W}$ and $\boldsymbol{H}$ with random non-negative values, and then continually update both matrices, element by element, according to,
\begin{equation}
    W_{ik} \leftarrow W_{ik}\frac{[\boldsymbol{W}^T\boldsymbol{H}]_{ik}}{[\boldsymbol{WHH}^T]_{ik}}, \;     H_{kj} \leftarrow H_{kj}\frac{[\boldsymbol{W}^T\boldsymbol{V}]_{kj}}{[\boldsymbol{W}^T\boldsymbol{WH}]_{kj}},
\end{equation}
in order to minimise Eq.~\ref{eq:euclid_distance}. In this work, we do not initialise the matrices randomly but instead use Non-negative Double Singular Value Decomposition \citep[][]{boutsidis2008svd} with zeros filled with small random values. Once a stable solution has been achieved, the matrix $\boldsymbol{H}$ can be used to reconstruct the now un-mixed sources. If the NMF is applied again to the same data, the random initial conditions can lead to different labelling of the components, but the actual reconstructed sources will be the same. Throughout this work, to conduct NMF, we use \texttt{sklearn.decomposition.NMF} \citep[][]{scikit-learn}, which has a loss function of the form  
\begin{align*}
    L(W,H) &= 0.5 \times ||\boldsymbol{V} - \boldsymbol{WH}||^2 \\
    &+ \alpha_W \times L_1 \times n \times \sum_{ij}||W_{ij}|| \\
    &+ \alpha_H \times L_1 \times m \times \sum_{ij}||H_{ij}|| \\
    &+ 0.5 \times \alpha_W \times (1 - L_1) \times n \times \sum_{ij} W_{ij}^2 \\
    &+ 0.5 \times \alpha_H \times (1 - L_1) \times m \times \sum_{ij} H_{ij}^2,
\end{align*}
where $\alpha_W, \alpha_H$ and $L_1$ are all regularisation parameters. Throughout this work, we choose to set regularisation parameters to zero. When a random grid search over all regularisation parameters was conducted, all sensible reconstruction error metrics favoured setting the regularisation terms to zero  (or very close to zero). Specifically, we ran a 5000 iteration random grid search with parameters $\alpha_H = \{10^{-3}, 10^0\}$, $\alpha_W = \{10^{-3}, 10^0\}$, both with 50 linearly spaced increments, and $LE1 = \{0.0, 1.0\}$ with 5 linearly spaced increments. 

\subsection{Data preparation}

In this work, we use NMF to find distinct chemical components across the entire range of energy $E$ and z-angular momentum $L_z$ space of the stellar halo. Therefore, to create the data matrix $\boldsymbol{V}$, we bin the data into regions with equal numbers of stars in energy and z-angular momentum space $(E, L_z)$. This results in a non-regular spacing of the bins across the $(E, L_z)$ plane. We then produce two $l \times l$ sized 2-d histogram image of [Al/Fe]--[Fe/H]  and [Mg/Fe]--[Fe/H] space for each $(E,L_z)$ bin, flatten each image vector into a $l^2$ length vector, and subsequently stack them together. Here $l$ is simply the number of pixels along one axis in the histogram image, where we take our fiducial value to be $l=50$. Therefore, in our setup, $n = 2l^2$ is the size of the two stacked flattened image vectors, $m=q^2$ is the number of bins in $(E, L_z)$ (we choose m to be a square number), and $p$ is a prescribed number of distinct chemical populations (components) that we want to find in each $(E,L_z)$ bin. 

\subsection{Re-application of components}\label{sec:repixel}

While it is sensible to find the original $m$ components in $(E,L_z)$ bins of equal signal-to-noise, this results in bins that are not regularly spaced, and thus sometimes difficult to interpret. Therefore, once the NMF components $\boldsymbol{H}$ have been found for a first pass (in equal SNR bins), we can apply these components to a new dataset $\boldsymbol{V}'$ with regular bin spacing and recover new weights $\boldsymbol{W}'$ for these new bins. In essence, we are now minimising some function of the a new error $L'$,
\begin{equation}\label{eq:new_error}
    L'  = 0.5 \times ||\boldsymbol{V}' - \boldsymbol{W}'\boldsymbol{H}||^2,
\end{equation}
with respect to $\boldsymbol{W}'$ using the same method as before.

Regularly spaced bins allow for a more easily interpretable space in which to study the distribution of the chemical components. Evidently, these bins will contain an unequal number of stars, so we want to ensure the number of data points in each bin is sufficiently large to make interesting conclusions. Moreover, we can apply the components to variables other than energy and angular momentum. If we bin the data by any variable (spherical radius, for example) we can then find the contribution of each of the pre-determined components as a function of that new variable. In essence, this re-defines our mixture space (i.e. our ``signals''), but uses the same ''sources'' as earlier, allowing us to map components as a function of a new variable.

\section{Halo Components}\label{sec:components}

In this section, we show the results of various NMF models that we applied to the APOGEE DR17 data, and the subsequent analysis. The NMF models vary in terms of their input data dimensions, input data content and, component number.

\subsection{Two component model}\label{sec:two_component_model}

We first present the results of considering an NMF model that is prescribed with $p=2$ components. This is physically motivated by the expectation that the stellar halo can broadly be split into an accreted population and an \textit{in-situ} population. We tested that the same components and weights were recovered with every repeated run of the NMF, with the same initial conditions.

\subsubsection{Chemically pure stellar halo}

\begin{figure*}
    \centering
    \includegraphics[width=0.95\textwidth]{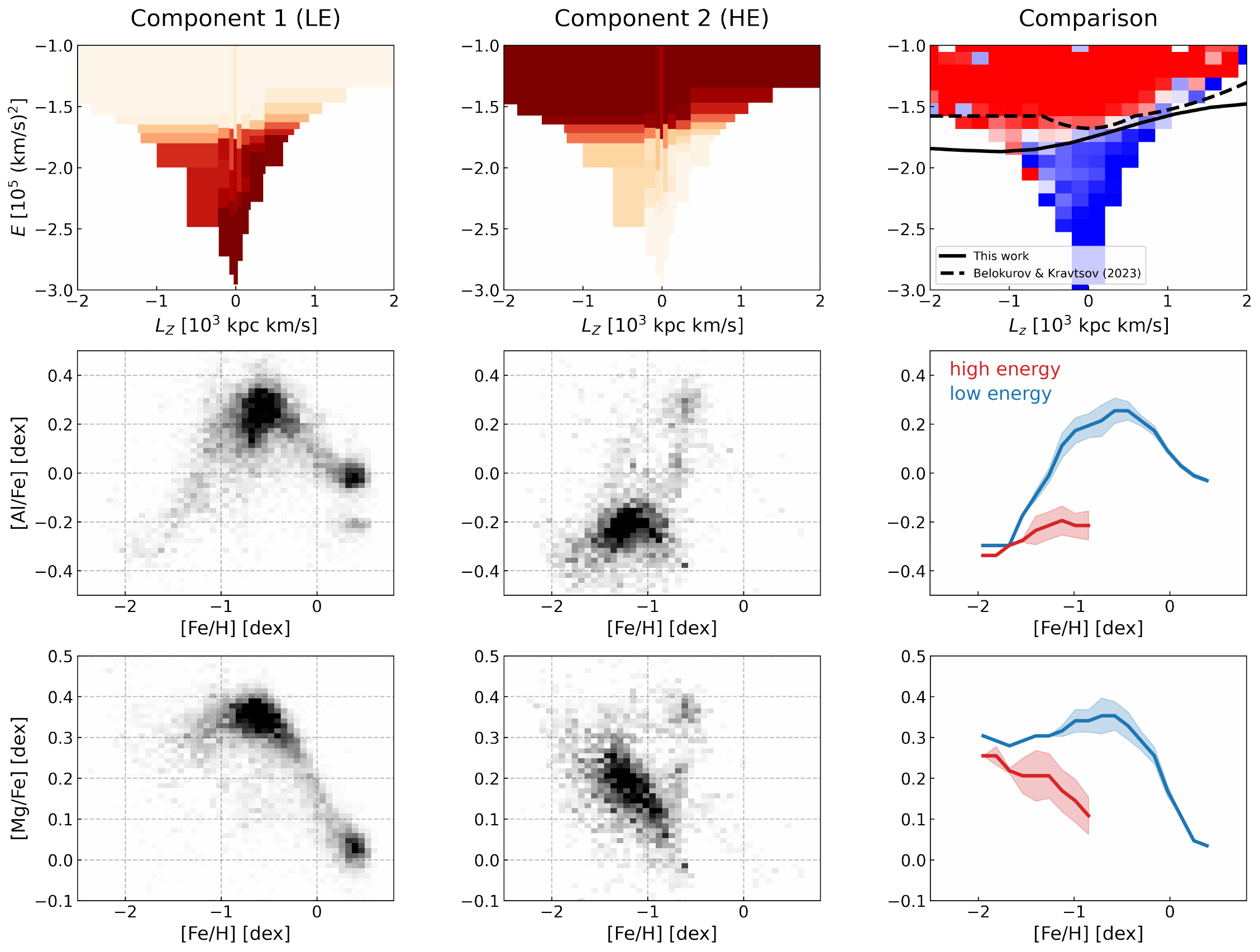}
    \caption{The resulting components from a 2-component NMF applied to the chemically pure stellar halo data. The tow row presents energy and z-angular momentum space, the middle row presents aluminium chemical information, whereas the bottom row presents magnesium chemical information. \textit{Left column:} The first (low energy) component that consist almost entirely of stars that would typically be categorised as \textit{in-situ}. While these stars cover a wide range of metallicities, they dominate at [Fe/H] > 1, and are mostly found above [Al/Fe] > 0. \textit{Middle colummn:} the second (high energy) component that consists of of stars that are normally categorised as accreted. This component is mostly found at [Fe/H] < -1 and [Al/Fe] < 0. \textit{Right column:} Comparison of the two components. The top-right panel shows a ratio of the two components, along with the boundary between the two components as a black solid line. Specifically, this boundary is obtained by finding (in each $L_z$ column) the point where the contribution of the accreted falls below $50\%$. The black dashed line shows the boundary between the accreted and \textit{in-situ} components from \citet[][]{Belokurov2023nitrogen}. The middle-right and bottom-right panels show median tracks (and median absolute deviation) of the components in the two left-most columns, for ease of comparison. The blue line shows the low-energy component, and the red line shows the high-energy component.}
    \label{fig:fig3}
\end{figure*}

Here, we present the results of conducting NMF on the APOGEE DR17 stellar halo data that is selected by making kinematic cuts, as illustrated by the middle column of Fig.~\ref{fig:fig1}. Firstly, we package this data appropriately into the data matrix $\boldsymbol{V}$ for $q=8$ and $l=50$, implying that the matrix $\boldsymbol{V}$ is of dimensions $64 \times 5000$, as outlined in the previous section. We first prescribe the NMF method to output $p=2$ components, which implies that the output matrices $\boldsymbol{W}$ and $\boldsymbol{H}$ will be of dimensions $64 \times 2$ and $2 \times 5000$ respectively. We explored other sensible ranges of $q$ and $l$ and found that the components changed very little.

We can see the result of this NMF model in Fig.~\ref{fig:fig3}. In the top-right and top-middle panel we show each of the 64 $(E,L_z)$ bins coloured by the the respective component weights in the $\boldsymbol{W}$ matrix, which have been scaled to take values between 0 and 1. Deep red indicates a strong contribution, whereas pale yellow indicates a weak contribution. In the middle row and bottom rows of the first two columns, we show the associated components produced from the unflattened $\boldsymbol{H}$ matrix, which are found mostly in the deep red regions of the top row. 

The top-right panel presents a re-application of the components to data that is binned by regular spacing in $(E,L_z)$ space. Specifically, we bin the stars in linearly spaced bins in energy, from $E = -3\times10^5$ (km/s)$^2$ to $-1\times10^5$ (km/s)$^2$ and linearly spaced bins in z-angular momentum from $L_z = -2\times10^3$ kpc km/s to $L_z = 2\times10^3$ kpc km/s. The 2-d histogram in the top-right panel shows the normalised (between 0 and 1) weights of the low energy component, for the case where we have used 20 linearly spaced bins in either direction. Therefore, deep blue indicates a value of 1 (low-energy component domination) and deep red indicates a value of 0 (high-energy component domination). Using this regularly spaced data, we can define a boundary between the high-energy and low-energy components, which roughly corresponds to a boundary between the accreted and \textit{in-situ} populations. To obtain a boundary, we find the $E$ value (for a given $L_z$ column) that corresponds to where the high-energy component first falls below a $50\%$ contribution. The black line in the top-right panel of Fig.~\ref{fig:fig3} shows this boundary, smoothed with a gaussian filter, for the case where we have used instead 10 linearly spaced bins to ensure a sufficient large number of stars per bin. We also plot the boundary between the accreted and \textit{in-situ} populations from \citet[][]{Belokurov2023nitrogen}. To find a functional form for our boundary, we fit a third degree polynomial (using \texttt{numpy.polyfit}) to the boundary found from the 10 bins. We obtain the following:

\begin{equation}
    E_{\rm b}(L_z) = -0.018 L_z^3 + 0.024 L_z^2 + 0.162L_z - 1.751
\end{equation}

for the range of angular momentum $|L_z| < 2\times10^3$ kpc km/s, where $L_z$ is in units of $10^3$ kpc km/s and $E_b$ is in units of $10^5$ (km/s)$^2$. This is of course only valid for the potential in \citet[][]{mcmillan2017mass}.

Overall, the boundary we find is close to that derived in the latter study, which was based on a simple but well-motivated threshold in [Al/Fe]. This is particularly true for the $L_z>0$ region. At the same time, at $L_z<0$ our boundary is at a lower energy, with the difference as large as $0.3-0.4\times 10^5\,\rm (km/s)^2$. While this boundary somewhat looks to coincide with the APOGEE selection function artifact, our NMF derived boundary is actually at a higher energy than the selection artifact. This is made more clear in Appendix.~\ref{sec:appendix1}, where the selection artifact is evidently at an energy of $E\sim-2.0\times10^5$~(km/s)$^2$ (see Fig.~\ref{fig:appendix1a} and Fig.~\ref{fig:appendix2}).

The middle-right and bottom-right panels show the median tracks of the components in [Al/Fe] -- [Fe/H] space and [Mg/Fe] -- [Fe/H] space, respectively. The blue line shows the track of the low-energy component and the red line shows the track of the high-energy component. The shaded areas around each median track illustrate the median absolute deviation (MAD).

Upon visual examination of these components, we can see that they occupy regions in chemical and kinematic space that one typically associates with the accreted and and \textit{in-situ} populations. Namely, there is a low-energy, high-metallicity, high-$\alpha$ component (leftmost column), and a high-energy, low-metallicity, low-$\alpha$ component (middle column). These broadly correspond to the \textit{in-situ} and accreted populations, respectively. Interestingly, however, the low-energy component has a tail that extends well below the typically accepted \textit{in-situ} limit of [Al/Fe]~$\sim 0$ dex and [Fe/H]~$\sim -1.1$ dex. Moreover, the high-energy component has an extended sequence at the higher metallicity end, that extends up towards the low-energy overdensity. This appendage appears to correspond with the recently discovered structure known as \textit{Eos} \citep[][]{myeong2022milky, matsuno2024distinct}. Whether or not these extended features correspond to some continuous sequence of star formation is beyond the scope of this work.

\subsubsection{Kinematically pure stellar halo}

Now we present the results of applying NMF to the stellar halo data that is selected by making chemical cuts (see the rightmost column of Fig.~\ref{fig:fig1}) and the red dashed line in Fig.~\ref{fig:fig2}). A kinematically pure dataset allows us to infer kinematic properties of the stellar halo without being subject to our previous kinematic cuts. As before, we package the data into a matrix $\boldsymbol{V}$ that is of dimensions $64 \times 5000$. The resulting components and weights are of the same dimensions as the chemically pure stellar halo. 

\begin{figure*}
    \centering
    \includegraphics[width=0.95\textwidth]{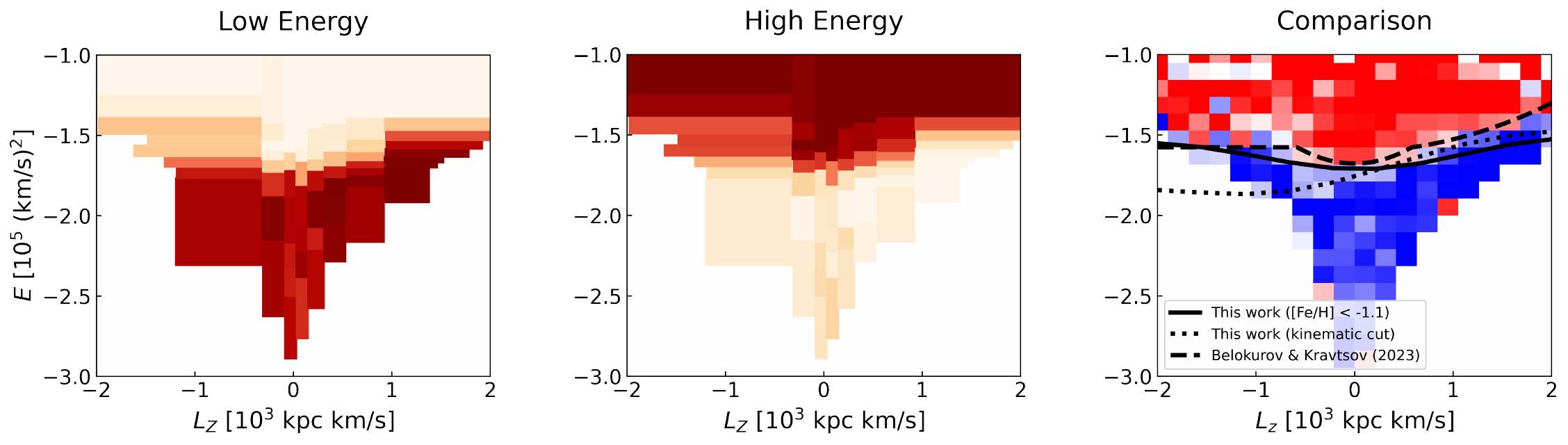}
    \caption{The same as the top row of Fig.~\ref{fig:fig3} but for the kinematically pure stellar halo, selected by choosing only stars with [Fe/H] < -1.1 dex. Note that the resulting boundary from this low-metallicity election more closely aligns with the result from \citep[][]{Belokurov2023nitrogen}.}
    \label{fig:fig4}
\end{figure*}

In Fig.~\ref{fig:fig4}, we present the resulting distribution of the components in $(E,L_z)$ space. As before, there is a clear split between high energy and low energy. However, the boundary between the two components is somewhat different, especially at negative $L_z$. Again, we find the point at which the high energy component's contribution falls below $50\%$, and define this as our boundary between the two components. The solid black line shows the boundary for the kinematically pure ([Fe/H] < -1.1 dex) stellar halo, whereas the dotted line shows the boundary for the chemically pure stellar halo. We again quantify this boundary using a polynomial fit, but find that a fourth order polynomial is required:

\begin{align}
    \tilde{E}_{\rm b}(L_z) = &-0.011L_z^4 + 0.002L_z^3 + 0.085L_z^2 \nonumber \\
    &- 0.003L_z - 1.711.
\end{align}

As before, this is valid for the range of angular momentum $|L_z| < 2\times10^3$ kpc km/s, where $L_z$ is in units of $10^3$ kpc km/s and $\tilde{E}_b$ is in units of $10^5$ (km/s)$^2$. We can see that this low-metallicity boundary is more aligned with the boundary estimated by \citet[][]{Belokurov2023nitrogen} than the ``chemically pure'' model. 

\subsection{Four component model}

With a new boundary that defines the approximate transition point between the accreted and \textit{in-situ} populations of the stellar halo, it seems natural to break up these populations further and investigate them in more detail. Naively, this could unveil sup-populations of accreted debris in the high energy component, or distinct low energy \textit{in-situ} populations. Therefore, we conduct separate NMF models on the high-energy population and low-energy population. For simplicities sake, we only consider breaking the HE and LE components up into two further $p=2$ component models, thereby resulting in a four component model overall. Given that the kinematically pure dataset already has a reduced number of stars, we only conduct this four component experiment on the chemically pure dataset. 

Using the functional form of the boundary, $E_{\rm b}(L_z)$, defined using the chemically pure sample (Fig.~\ref{fig:fig3}) we create distinct datasets for the high-energy and low-energy stars. Given the decrease in the number of stars in each dataset, we package the data differently than before, using only $m=25$ ($q=5$) mixtures. However, we keep the same number of bins in chemical space, $l=50$. This results in an input matrix $\boldsymbol{V}$ of dimensions $25 \times 5000$, a weight matrix $\boldsymbol{W}$ of dimensions $25\times2$, and a component matrix $\boldsymbol{H}$ of dimensions $2\times5000$.

\begin{figure*}
    \centering
    \includegraphics[width=0.99\textwidth]{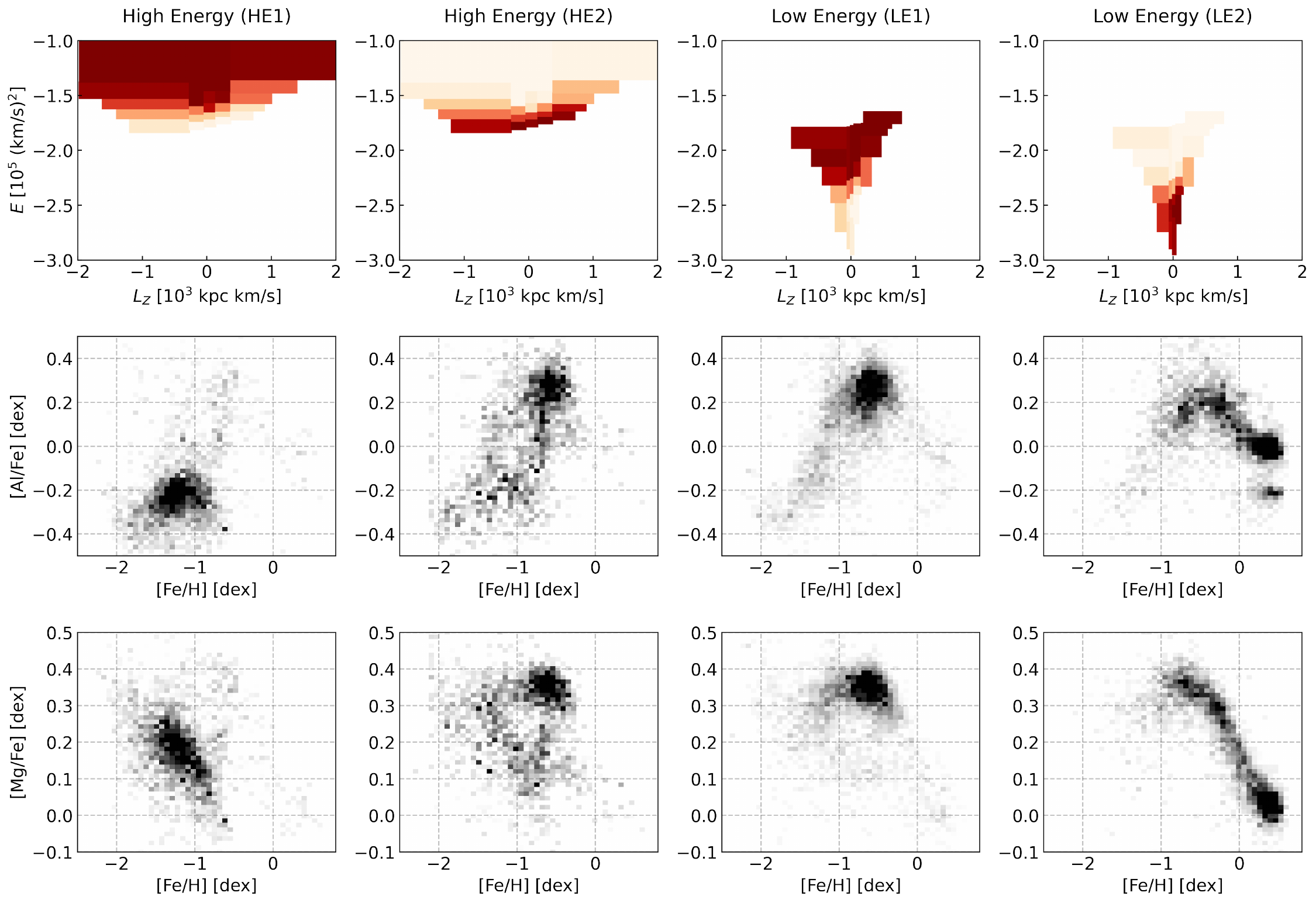}
    \caption{The resulting components from applying NMF to the two datasets that come from separating the data into a high-energy sub-dataset and low-energy sub-dataset, based on the boundary presented in Fig.~\ref{fig:fig3}. We present the components in order of decreasing energy from left to right, and refer to them as HE1, HE2, LE1 and LE2. Component HE1 presents a reasonably pure sample of accreted stars, concentrated around [Fe/H] < -1. Component HE2, however, more clearly highlights the distinct connections between the high [Al/Fe] and low [Al/Fe] sequence, seen at around [Al/Fe] = 0. }
    \label{fig:fig5}
\end{figure*}

We present the result of these two NMF models in Fig.~\ref{fig:fig5}, where we show the different components in order of decreasing energy from left to right. In that order we label these components HE1, HE2, LE1, LE2. Component HE1 is dominated by the low-[Fe/H], low-[Al/Fe] region typically associated with the GSE. This component appears to be a reasonably pure sample of accreted stars, concentrated around [Fe/H] < -1.0 dex and [Al/Fe] < -0.2 dex. Component HE2 has some overlap with HE1 at [Fe/H] < -1.0 dex and [Al/Fe] < -0.2 dex, as well as some overlap with LE1 at [Al/Fe] > 0.2 dex. While HE2 contains content from these tail ends of the [Al/Fe] distribution, it highlights some interesting branches joining these two extremes, around [Al/Fe]~$\sim 0.0$ dex. The higher metallicity branch ([Fe/H] > -1.0) is what has been referred to as \textit{Eos}, whereas the lower metallicity branch ([Fe/H] < -1.0) is what has been referred to as either Aurora or Thamnos. Components LE1 and LE2 show how the low-energy region of the halo is split into two main overdensities centred on [Fe/H]~$\sim -0.6$ dex and [Fe/H]~$\sim 0.4$ dex. While HE2 highlights the \textit{Eos} branch more clearly than LE1, is it likely they are part of the same fundamental chemical component and are artificially broken in two by our initial separation into high and low energy. If we follow the sequence in component LE2, interestingly, it does not reach the same peak value of [Al/Fe] = 0.4 dex as the main overdensity in component LE1. Instead, it peaks at just above [Al/Fe] = 0.2 dex. Likewise, the \textit{Eos} branch at [Al/Fe]~$\sim 0.0$ and (Fe/H] > -1.0 does seem to connect directly to the LE1 overdensity at [Al/Fe]~$\sim 0.2$. This could suggest that we have uncovered two distinct sequences of evolution, and not just split the low-energy component in half. Component LE1 is dominated by the high [Al/Fe] disk stars, often referred to as the Splash, whereas component LE2 highlights the high metallicity thick disk. Component LE1 shows significant overlap with HE2, as well as some with HE1, most notably at [Fe/H] < -1.0 dex and [Al/Fe] < -0.2 dex. The Mg bifurcation of \textit{Eos} is quite visible in the [Mg/Fe] -- [Fe/H] space of component HE2, and less-so in LE1. Note also the small tail at [Fe/H]$\sim -0.5$ dex and [Al/Fe]$\sim 0.1$ dex which looks to be some overlap of \textit{Eos} between the two components. We note that component HE2 and LE1 share common features: Splash and \textit{Aurora} (discussed in more detail later). It is possible that, in LE2, we have picked up some inner Galactic stars from the bulge and/or bar, whose chemical evolution is known to be distinct from the inner stellar halo.

\subsection{One-dimensional chemical distributions}

\begin{figure}
    \centering
    \includegraphics[width=0.9\columnwidth]{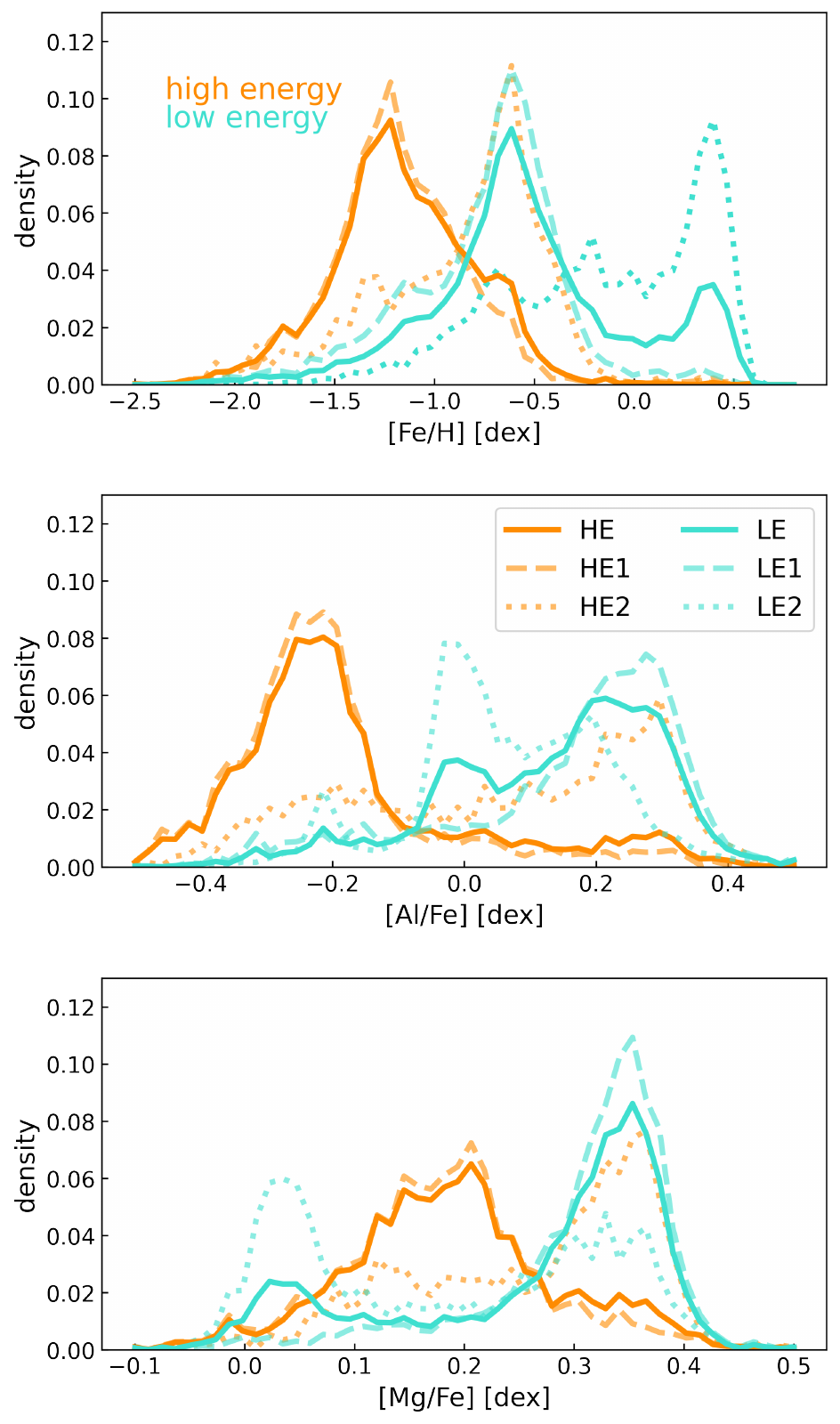}
    \caption{One dimensional density distributions of the chemical abundances in the components from the two-component (Fig.~\ref{fig:fig3}) and four component (Fig.~\ref{fig:fig5}) NMF models, with chemically pure input data. The orange lines present the distributions of the high energy components, whereas the cyan lines present the distributions of the low energy components. The solid lines show the 2-component distributions, whereas the dashed and dotted lines show the 4-component distributions.}
    \label{fig:fig6}
\end{figure}

To more quantitatively assess the components resulting from both the 2-component and 4-component chemically pure stellar halo, we plot the 1-d density distributions of each chemical abundance that is used in the NMF model in Fig.~\ref{fig:fig6}. From top to bottom, we show the distribution of [Fe/H], [Al/Fe], and [Mg/Fe]. The solid lines show the distribution of the 2-component model (chemically pure), and the dashed and dotted lines show the distributions of the 4-component models. Orange lines represent the distributions of the high-energy components, and cyan lines represent the distributions of the low-energy components. Dashed lines represents HE1 and LE1, whereas dotted lines represent HE2 and LE2.

The solid lines shows that the high-energy population peaks at lower values of [Fe/H], [Al/Fe] and [Mg/Fe] than the low-energy population. In Table.~\ref{tab:table1}, we list the 10th, 50th and 90th percentiles of chemical information from the 2-component and 4-component NMF models that take the chemically pure stellar halo information as input data. We do not provide information from the kinematically pure stellar halo, since cuts are applied to the chemistry before they are used in the model. Table.~\ref{tab:table1} shows that the stellar halo is dominated by the low-energy (approximately \textit{in-situ}) component, at over $66\%$, while the high-energy (approximately accreted) component makes up only around $34\%$. These results match up closely with what is found for globular clusters by \citet[][]{belokurov2024insitu} and \citet[][]{chen2024galaxy}. We point out in Appendix.~\ref{sec:appendix1} that removing the inner 5 kpc from the chemically pure sample changes these contributions to $47\%$ for the \textit{in-situ} population and to $53\%$ for the accreted population.

The major peaks of components HE and LE (solid lines) line-up exactly with HE1 and LE1 (dashed lines) for all abundances. However, we can see that the smaller peaks are flattened out with the introduction of the 4-component model. For example, see the small orange bump in HE at [Fe/H]~$\sim -0.7$ dex, which becomes flattened in HE1, yet is the primary peak in HE2. This introduction of 4-components has allowed us to further break down the halo into more well-defined clumps. The small orange bump comes from \textit{Eos} and some contamination from the Splash, which have mostly been removed in HE1. Likewise, compare the cyan lines at [Mg/Fe]~$\sim 0.05$ and [Al/Fe]~$\sim 0.00$ dex. We can see here that the 4-component model has cleaned up the distributions such that LE1 and LE2 are more pure samples of their respective two populations.

We can see very clearly from the middle panel of Fig.~\ref{fig:fig6}, that component LE1 and LE2 fall of at different maximum values of [Al/Fe]. If we compare the dashed and dotted cyan lines, we can see that the dotted line (LE2) falls off beyond [Al/Fe]~$\sim 0.2$ but the dashed line (LE1) falls of beyond [Al/Fe]~$\sim 0.3$ -- $0.4$. This suggests that we have picked up unique tracks in LE1 and LE2, and the turn-around point of LE2 is not simply a result of contamination from the other component. This could suggest we have identified two unique sequences of star formation. Strangely, however, this misalignment is not seen in [Mg/Fe]. We can see that in the bottom panel, both dotted and dashed cyan lines fall off beyond [Mg/Fe]~$\sim 0.4$. However, it is likely that component LE2 is populated with stars from the bar and/or bulge, and this unique chemical sequence is illustrative of that region of the MW.

\begin{table}
\caption{Percentiles of chemical abundances, and contribution, of the components resulting from the two component and four component NMF models that used chemically pure stellar halo input data.}
\label{tab:table1}
\scalebox{0.9}{%
\begin{tabular}{llllll}
\hline
\multicolumn{1}{|l|}{Component} &
  \multicolumn{1}{l|}{\%} &
  \multicolumn{1}{l|}{Percentile} &
  \multicolumn{1}{l|}{{[}Fe/H{]}} &
  \multicolumn{1}{l|}{{[}Al/Fe{]}} &
  \multicolumn{1}{l|}{{[}Mg/Fe{]}} \\ \hline \hline
\multirow{3}{*}{HE (high-energy)} & \multirow{3}{*}{$33.53$} & 10  & $-1.62$ & $-0.36$ & $0.07$ \\
                                  &                          & 50 (median) & $-1.15$ & $-0.21$ & $0.18$ \\
                                  &                          & 90  & $-0.68$ & $ 0.13$ & $0.33$ \\ \hline
\multirow{3}{*}{LE (low-energy)}  & \multirow{3}{*}{$66.47$} & 10  & $-1.15$ & $-0.09$ & $0.05$ \\
                                  &                          & 50 (median) & $-0.55$ & $ 0.17$ & $0.32$ \\
                                  &                          & 90  & $ 0.33$ & $ 0.32$ & $0.38$ \\ \hline \hline
\multirow{3}{*}{HE1}              & \multirow{3}{*}{$58.32$} & 10  & $-1.62$ & $-0.36$ & $0.07$ \\
                                  &                          & 50 (median) & $-1.22$ & $-0.23$ & $0.18$ \\
                                  &                          & 90  & $-1.75$ & $ 0.01$ & $0.29$ \\ \hline
\multirow{3}{*}{HE2}              & \multirow{3}{*}{$41.68$} & 10  & $-1.49$ & $-0.28$ & $0.11$ \\
                                  &                          & 50 (median) & $-0.75$ & $ 0.09$ & $0.30$ \\
                                  &                          & 90  & $-0.41$ & $ 0.32$ & $0.38$ \\ \hline \hline
\multirow{3}{*}{LE1}              & \multirow{3}{*}{$42.54$} & 10  & $-1.29$ & $-0.15$ & $0.18$ \\
                                  &                          & 50 (median) & $-0.68$ & $ 0.21$ & $0.33$ \\
                                  &                          & 90  & $-0.34$ & $ 0.34$ & $0.38$ \\ \hline
\multirow{3}{*}{LE2}              & \multirow{3}{*}{$57.46$} & 10  & $-0.88$ & $-0.11$ & $0.01$ \\
                                  &                          & 50 (median) & $-0.08$ & $ 0.05$ & $0.18$ \\
                                  &                          & 90  & $ 0.46$ & $ 0.26$ & $0.37$ \\ \hline
\end{tabular}%
}
\end{table}

\subsection{Eccentricity distribution}

Using the NMF components resulting from the kinematically pure stellar halo data (made with a metallicity cut of [Fe/H]~$< -1.1$ dex), we can explore some kinematic properties of the halo. One of the most interesting properties to examine is eccentricity, which can give insight into the formation history of the accreted population. Given that eccentricity is not something we provide directly to the NMF model, we must find some other way to obtain this information from the components. 

\begin{figure}
    \centering
    \includegraphics[width=0.95\columnwidth]{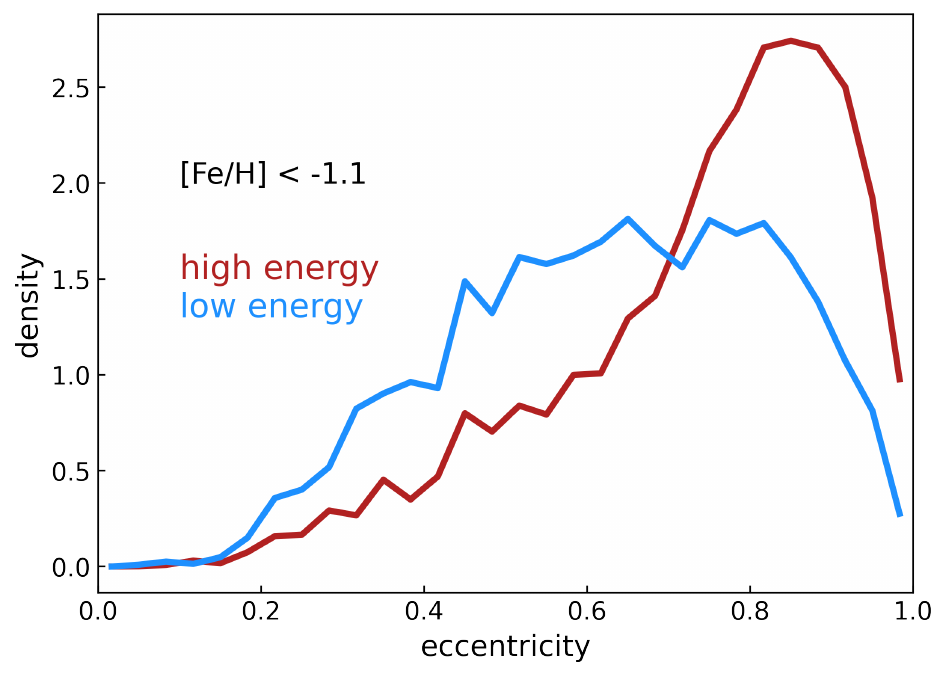}
    \caption{Eccentricity distribution of the kinematically pure ([Fe/H] < -1.1 dex) stellar halo dataset. The high-energy component distribution is shown by a red line, whereas the low-energy distribution is shown as a blue line. The high energy component peaks at around $e \simeq 0.85$, whereas the low energy component peaks at around $e \simeq 0.6$, and has a higher proportion of stars at lower eccentricity.}
    \label{fig:eccentricity}
\end{figure}

In each of the 64 $(E,L_z)$ bins, there is an associated distribution of eccentricities. To find the overall distribution of eccentricities for each component, we can use the components as a mask over the bins. Specifically, for a given component, we choose to consider every bin that contributes more than $10\%$ of the maximum bin value. This gives us a subset of bins the $m=64$, in which there are distributions of eccentricities $f_m(e)$ in each bin $m$. Moreover, each $(E,L_z)$ bin has an associated weight from the $\boldsymbol{W}$ matrix. We multiply the weight of each bin, by the distribution in each bin, to find a total distribution of eccentricities $F_p(e)$ for a given component $p$:
\begin{equation}
    F_p(e) = \sum_m W_{mp} f_m(e),
\end{equation}
where in this case $p=\{0,1\}$. In Fig.~\ref{fig:eccentricity}, we plot the resulting total distributions of eccentricities for both components. We can see that the high-energy components peaks at a value of approximately $e~\sim~0.85$, while the low-energy component peaks at around $e~\sim~0.60$. While the high-energy component peaks at a higher eccentricity, it still has a wide range of eccentricities. Distributions like this could be used to infer the mass of the GSE, based on the work by \citet[][]{amarante2022gastro}. In this paper the authors present smoothed particle hydrodynamics plus {\it N}-body  simulations of a single accretion event in a MW-like galaxy. Crucially, they show a link between the eccentricity distribution of their accreted debris and the initial total mass of the dwarf galaxy (e.g. see Fig.~6). Unfortunately, our sample is likely not a pure GSE sample, and so we can not make too many inferences based on this eccentricity distribution.

\section{Application of components}\label{sec:application}

In this section we take the components found from the NMF models described in Sec.~\ref{sec:components} and apply them to different datasets, as described in Sec.~\ref{sec:repixel}. Crucially, we vary the number of histogram bins $l$ to assess the robustness of our results. Specifically, we allow values of $l = \{10, 60\}$. The errorbars on the plots in this section reflect the variation in results from these different set-ups.

\subsection{Contribution as a function of energy}\label{sec:contribution_as_a_function_of_energy}

\begin{figure}
    \centering
    \includegraphics[width=0.95\columnwidth]{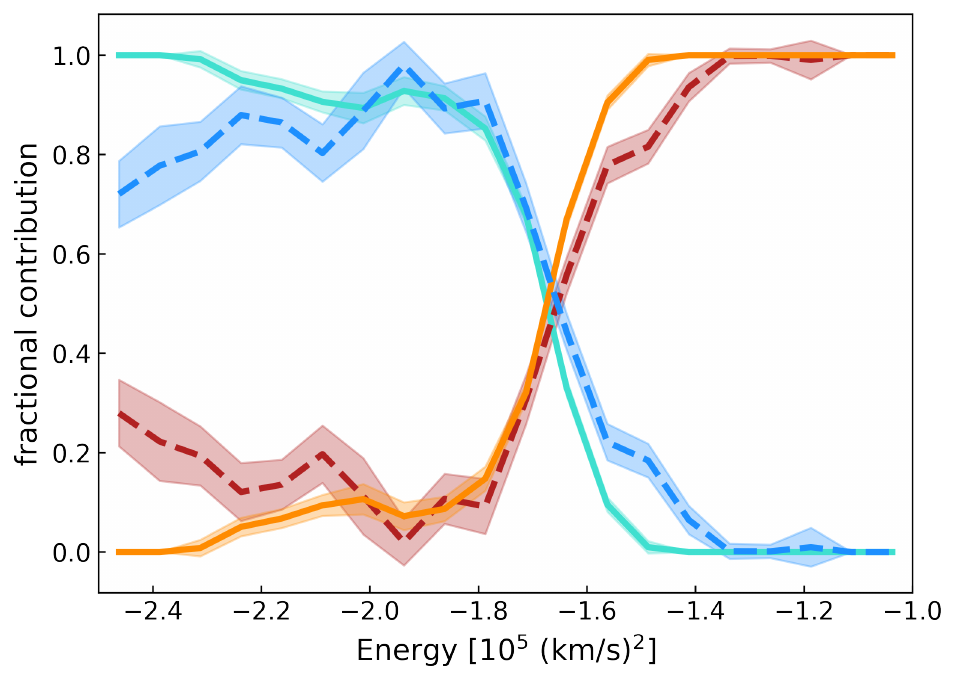}
    \caption{Fractional contributions of the $p=2$ NMF model components as a function of energy. The orange and red lines show the high-energy components, whereas the cyan and blue lines show the low-energy components. The solid lines show the contribution of the chemically pure stellar halo, whereas the dashed lines show the contributions of the kinematically pure stellar halo. By fitting a spline to the mean values, we find the fractional contribution as a function of energy. For the chemically pure sample, we find that the contribution of the accreted component is 0.5 at $E = -1.67\times10^5$ (km/s)$^2$, is 0.9 at $E = -1.56\times10^5$ (km/s)$^2$, and is 0.1 at $E = -1.84\times10^5$ (km/s)$^2$. For the kinematically pure data, with the fractional contribution of the accreted component is 0.5 at $E = -1.65\times10^5$ (km/s)$^2$, and is 0.9 at $E = -1.43\times10^5$ (km/s)$^2$.}
    \label{fig:fig7a}
\end{figure}

First, we re-bin both datasets (pure kinematic and pure chemical) based on just energy. Specifically, we create 20 regularly spaced bins from $E=-2.5\times10^5$ to $E=-1.0\times10^5$ (km/s)$^2$ to examine the exact contribution of the two components as a function of energy. This allows us to estimate, for a given energy, how much accreted vs \textit{in-situ} debris there is. This obviously assumes the high-energy and low-energy components correspond to the accreted and \textit{in-situ}, respectively. We only look at the contributions as a function of energy for the 2-component models, given that each component of the 4-component model has such a limited range in energy. We show the result of this in Fig.~\ref{fig:fig7a}. The orange and red lines show the high-energy components, and the cyan and blue lines show the low-energy components. The dashed lines show the components produced from the kinematically pure halo data, and the solid lines show the components produced form the chemically pure halo data. The lines show the mean values from the all the different set-ups, whereas the shaded area around the line shows the $1\sigma$ values. 

By fitting a spline to the mean values, we can find the energy at which we flip from \textit{in-situ} domination to accreted domination. For the chemically pure sample, we find that this occurs at $E = -1.67\times10^5$ (km/s)$^2$. The contribution of the accreted component is 0.9 at $E = -1.56\times10^5$ (km/s)$^2$, and is 0.1 at $E = -1.84\times10^5$ (km/s)$^2$. The mean value of the boundary provided by \citet[][]{Belokurov2023nitrogen} is $E=-1.56\times10^5$ (km/s)$^2$. We find that, at this energy level, there is a $10\%$ contribution from the low-energy component, implying that there may be approximately $10\%$ \textit{in-situ} debris below this typically accepted boundary. 

The trend is much the same for the kinematically pure data, with the $50\%$ cross-over being at $E = -1.65\times10^5$ (km/s)$^2$, and the $90\%$ high-energy contribution at $E = -1.43\times10^5$ (km/s)$^2$. The only major difference is at lower energies, which may simply result from a lack of data.

\subsection{Contribution as a function of distance}\label{sec:contribution_as_a_function_of_distance}

Next we explore the contribution of the resulting NMF components as a function of various distances. Specifically, we look at galactocentric spherical radius $r$, height $z$, and galactocentric cylindrical radius $R$. We do this for the 2-component model (both kinematically pure and chemically pure datasets), as well as the 4-component model. We re-bin both datasets by linearly spaced bins in the distances mentioned earlier. We create 10 bins for each distance. Specifically, $r$ ranges from 2 to 14 kpc, $z$ ranges from 0 to 10 kpc, and $R$ ranges from 2 to 14 kpc.

\subsubsection{Two component contributions}

\begin{figure*}
    \centering
    \includegraphics[width=0.99\textwidth]{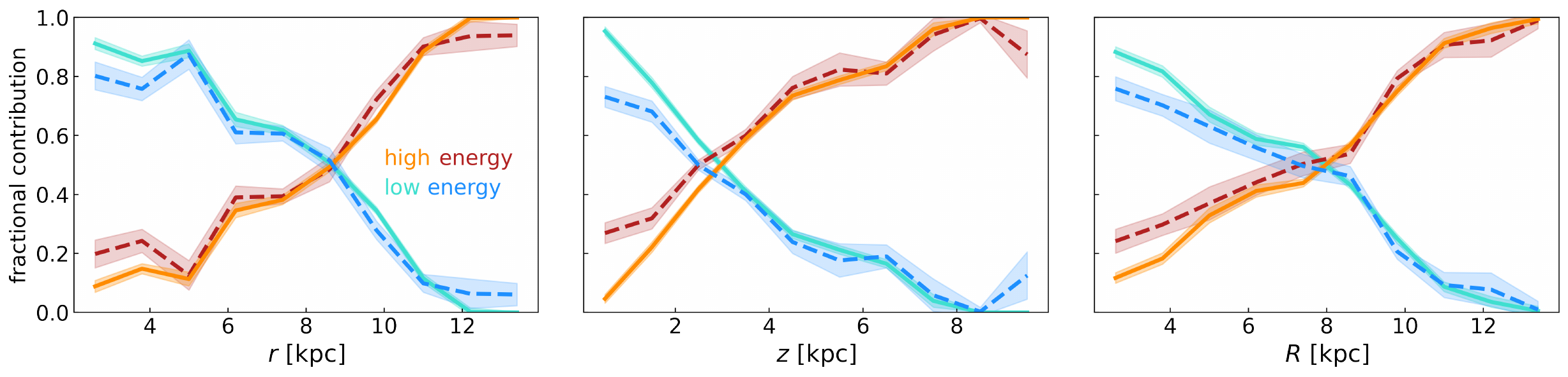}
    \caption{Fractional contribution of the $p=2$ NMF model as a function of galactocentric spherical radius ($r$), galactocentric height ($z$) and galactocentric cylindrical radius ($R$). The orange and red lines indicate the high-energy components, whereas the cyan and blue lines indicate the low-energy components. The solid lines show the contribution of the chemically pure stellar halo, whereas the dashed lines show the contribution of the kinematically pure stellar halo. While the spherical radius is similar in both, the kinematically pure stellar halo has a transition point at slightly lower height and cylindrical radius. We fit a spline to quantify the fractional contribution as a function of these distances and find that there is a $50\%$ contribution to each component at $(r,z,R) = (8.7, 3.0, 8.1)$ kpc for the chemically pure halo, and $(r,z,R) = (8.7, 2.5, 7.1)$ kpc for the kinematically pure halo.}
    \label{fig:fig7}
\end{figure*}

In Fig.~\ref{fig:fig7} we present the fractional contribution of the chemically pure halo components (solid lines) alongside the contribution of the kinematically pure halo components (dashed lines). The orange and red lines show the contribution of the high-energy components, whereas the cyan and blue lines show the contribution of the low-energy components. The lines show the mean values from the all the different set-ups, whereas the shaded area around the line shows the $1\sigma$ values. 

Like before, we fit a spline to the mean values to find the point at which there is a change from high-energy domination to low-energy domination. By doing this, we find that there is a $50\%$ contribution to each component at $(r,z,R) = (8.7, 3.0, 8.1)$ kpc for the chemically pure halo, and $(r,z,R) = (8.7, 2.5, 7.1)$ kpc for the kinematically pure halo. The transition from \textit{in-situ} domination to accretion domination therefore occurs just half a kpc beyond the solar radius, which matches what was found by \citet[][]{belokurov2024insitu}. For both data types, the galactocentric spherial radius transition point is very similar, but the other values are slightly lower for the kinematically pure halo.

\subsubsection{Four component contributions}

Since the 4-components were generated from two separate NMF models, we had to create a new singular 4-component NMF model to obtain the correct fractional contributions. Specifically, we apply the four components obtained from the two separate NMF models (presented in Fig.~\ref{fig:fig5}), to the distance-binned datasets described earlier. In Fig.~\ref{fig:fig7a} we present the fractional contribution of these four components. 

As expected, for all distances $(r,z,R)$, the high energy component dominates at larger distances. This seems sensible given that we expect to find more accreted debris than \textit{in-situ} debris in the outer halo. Components HE2 and LE1 mostly trend with each other, but the lowest energy component (LE2), shows a clear increase at $(r,z,R) \lesssim (4,2,4)$ kpc, and essentially no contribution beyond that. Therefore, this metal-rich low-energy component is found only at the inner most region of the Galaxy, which suggests that we may have picked up stars from the bar and/or bulge region.

Since there are distinct differences in the [Al/Fe], [Mg/Fe] and [Fe/H] between the four components, these distance fractional contributions allow us to broadly map out the gradients of these abundances as a function of radius and height across the stellar halo. We can say that the very inner halo (i.e. $(r,z,R) \lesssim (4,2,4)$ kpc) is [Fe/H]-rich, [Mg/Fe]-poor, and [Al/Fe]-intermediate. While the outer halo (i.e. $(r,z,R) \gtrsim (9,4,9)$ kpc) is [Fe/H]-poor, [Mg/Fe]-intermediate and [Al/Fe]-poor.

\begin{figure*}
    \centering
    \includegraphics[width=0.99\textwidth]{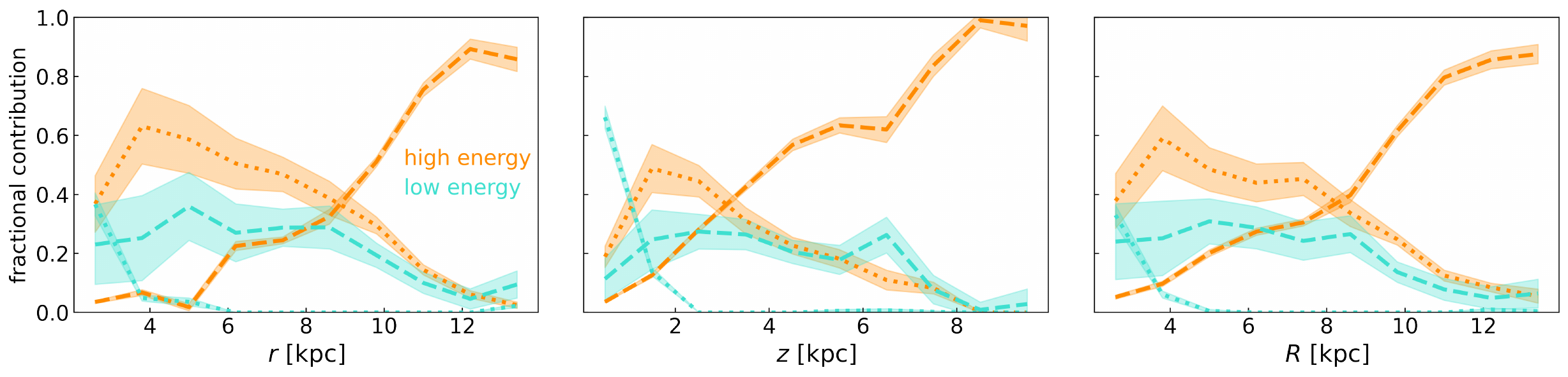}
    \caption{Fractional contribution of the $p=4$ NMF model as a function of galactocentric spherical radius ($r$), galactocentric height ($z$) and galactocentric cylindrical radius ($R$). The orange lines indicate the high-energy components, whereas the cyan lines indicate the low-energy components. Dashed lines indicate the higher energy components (i.e. HE1 and LE1), whereas the dotted lines indicate the low energy components (i.e. HE2 and LE2). Components HE2 and LE1 follow very similar trends, which is to be expected given their similar appearances in chemical space (see Fig.~\ref{fig:fig5}). Component LE1 dominates at inner height and radii, whereas component HE1 dominates at outer height and radii. By examination of the chemical space of the components, we can see that the very inner halo (i.e. $(r,z,R) \lesssim (4,2,4)$ kpc) is [Fe/H]-rich, [Mg/Fe]-poor, and [Al/Fe]-intermediate while the outer halo (i.e. $(r,z,R) \gtrsim (9,4,9)$ kpc) is [Fe/H]-poor, [Mg/Fe]-intermediate and [Al/Fe]-poor.}
    \label{fig:fig8}
\end{figure*}

\subsection{Comparison with simulations}

To better contextualise some of our results for the two component decomposition of stellar halo we compare the fractional contributions of the kinematically pure ([Fe/H]~$< -1.1$) stellar halo dataset with the galaxy formation simulations of galaxies in $z=0$ halos of mass $M_{\rm 200c}\approx 10^{12}\, M_\odot$ from the FIRE-2 suite \citep[][]{Wetzel2023}. Specifically, we present results of the simulations of MW-sized objects {\tt m12b, m12c, m12f, m12i, m12m, m12r}, and {\tt m12w}. 

\begin{figure*}

\includegraphics[width=1\linewidth]{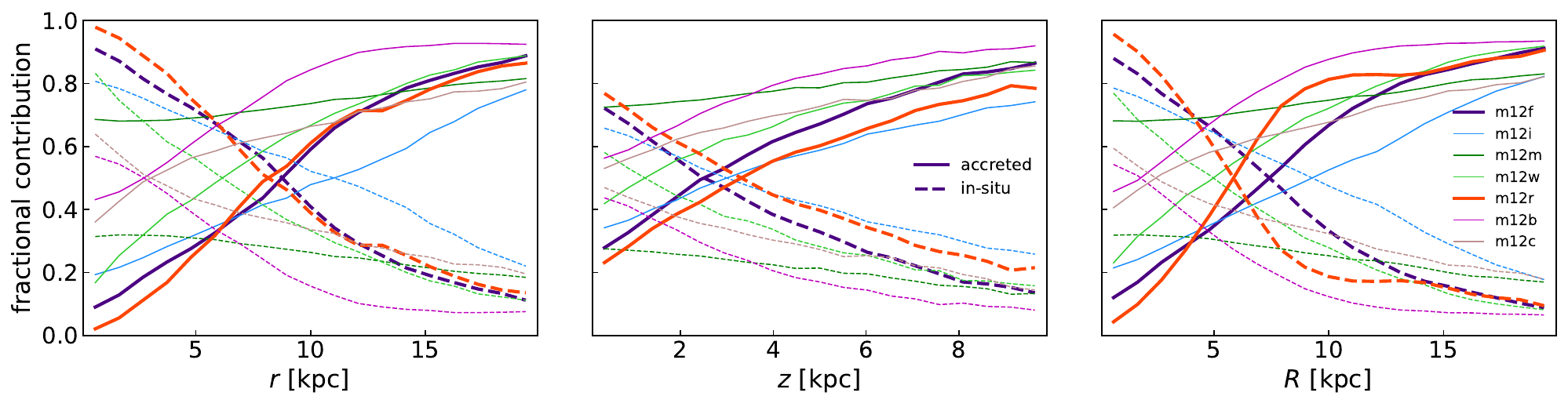}

\caption[]{Fractional contribution of the actual \textit{in-situ} and accreted stellar particles of metallicity ${\rm [Fe/H]}<-1.1$ in the FIRE-2 galaxy formation simulations of MW-sized galaxies (individual objects are labeled as {\tt m12f, m12i, m12m, m12w, m12r, m12b, m12c}), as a function of galactocentric spherical radius $(r)$, height $(z)$, and cylindrical radius $(R)$. The thicker lines show the two objects {\tt m12f} and {\tt m12r} with the fractional contribution profiles closest to those of the identified two components of the Milky Way halo shown  in Fig~\ref{fig:fig7}. }
\label{fig:fire_sims}
\end{figure*}

Figure~\ref{fig:fire_sims} shows the fractional mass contributions of the actual \textit{in-situ} and accreted stellar particles of metallicity ${\rm [Fe/H]}<-1.1$ in simulations as a function of galactocentric spherical radius $(r)$, height $(z)$, and cylindrical radius $(R)$. The corresponding distributions in the $E-L_z$ plane of the accreted and in-situ stars in these simulations can be found in Figure B1 in the Appendix B of \citet{belokurov2024insitu}. 

The figure shows that MW-sized galaxies in the simulations exhibit a broad range of the fractional contribution profiles, although all simulations generally show increasing contribution of accreted material with increasing distance from the galaxy center and the midplane of the disk. The specific form of these profiles likely reflects specifics of the evolution of each object and its accretion history. The model objects that have the profiles closest to those of the Milky Way can be considered to be MW analogues. In this sample, two objects {\tt m12f} and {\tt m12r} (shown by thicker lines in Fig.~\ref{fig:fire_sims}) have the profiles closest to those measured for the Milky Way in this analysis. Interestingly, as discussed in \citet[][]{belokurov2024insitu}, stellar particles in {\rm m12f} and {\tt m12r} have also the age-metallicity distribution closest to that of the Milky Way globular clusters. 

On the other hand, the stellar halo of {\tt m12m} is dominated by accreted material at most distances even in the central regions of the galaxy. Many of the other FIRE-2 hosts have cross-over points (i.e., where they transition form \textit{in-situ} dominated to accreted domination) much closer to the galactic centre than in the observed MW.
All of the simulated galaxies also have more extended distribution of {\it in-situ} stars than the Milky Way, where contribution of such stars to the stellar halo becomes negligible at distances $>12$ kpc.

\section{Component Sub-populations}\label{sec:deepdive}

In this section we take a deeper dive into some of the smaller features that appears in our components, and try to understand what they imply for the formation history of the MW. We label some of these features in Fig.~\ref{fig:fig12}.

\subsection{Eos}

\begin{figure}
    \centering
    \includegraphics[width=0.9\columnwidth]{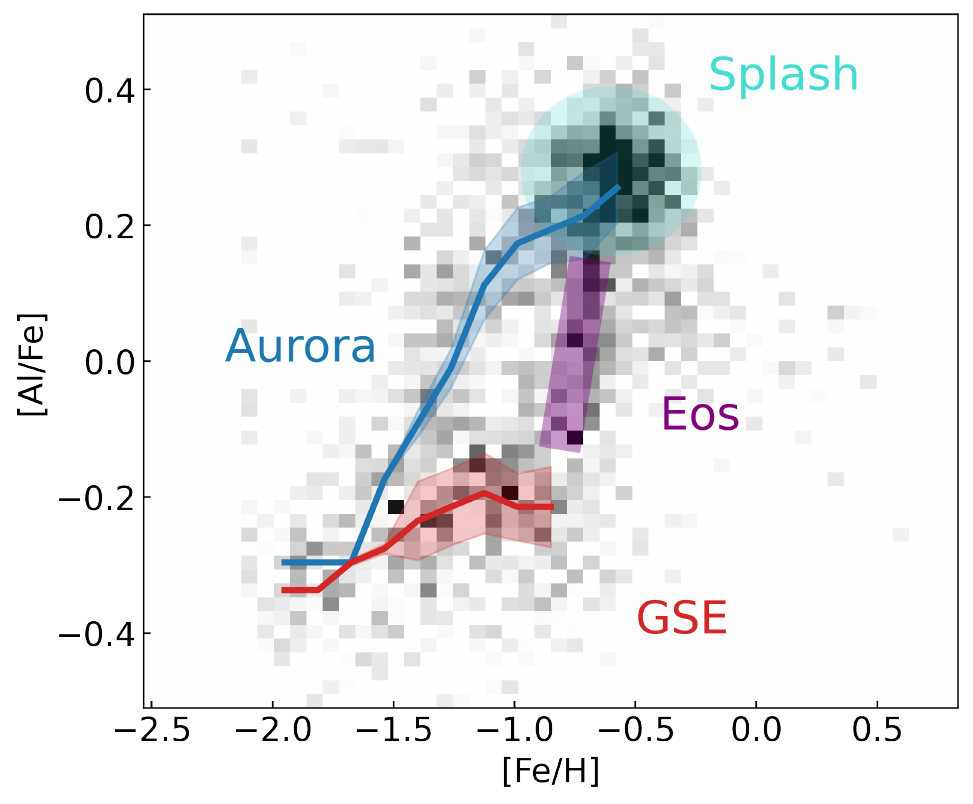}
    \caption{Zoom-in of the [Al/Fe] -- [Fe/H] space of component HE2 in Fig.~\ref{fig:fig5}, where the sub-populations have been labelled. Specifically, we identify the accreted GSE (red line), \textit{Eos} (purple rectangle), \textit{Aurora} (blue line) and the Splash (turquoise oval).}
    \label{fig:fig12}
\end{figure}

The most striking feature in in component HE2 in the middle panel of Fig.~\ref{fig:fig5} (and in Fig.~\ref{fig:fig12}), is the branch between the high metallicity GSE and the Splash, which we have labelled as \textit{Eos}. This labelling comes from the resemblance that this feature has with the structure of the same name in the work by \citet[][]{myeong2022milky}, which is described as being kinematically similar to the accreted halo, yet chemically similar to the disk. With these unusual properties its exact nature is unknown, but it is theorised to be a population of stars born \textit{in-situ} from gas polluted by the GSE merger. 

\begin{figure*}
    \centering
    \includegraphics[width=0.95\textwidth]{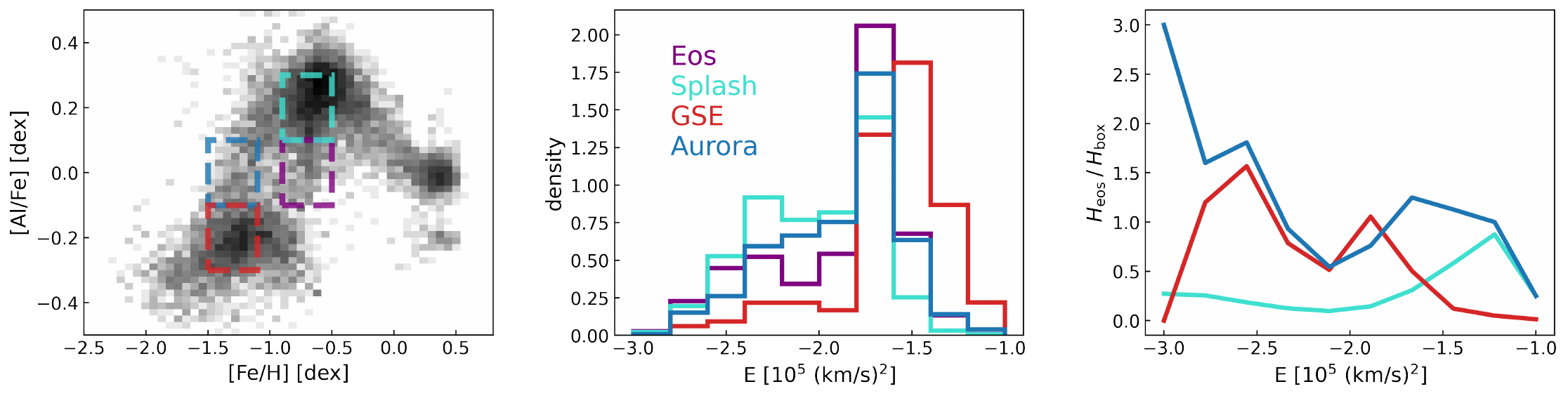}
    \caption{Comparison of the energy distribution of \textit{Eos} with other regions of the stellar halo. The leftmost panel shows the selections made for \textit{Eos} (purple), \textit{Aurora} (blue), GSE (red) and the Splash (turquoise). The middle panel shows a 10 bin 1-d density histogram of the energies of the stars in these bins. The right panel shows the ratio of the \textit{Eos} histogram to the other three selections. We can see that the energy distribution of \textit{Eos} is most similar to the distribution of the Splash.}
    \label{fig:fig9}
\end{figure*}

To shed light on this conundrum, in a way that is agnostic to the APOGEE selection effects, we compare the energy distributions of \textit{Eos} and the other sub-components. In Fig.~\ref{fig:fig9}, we show four selections in [Al/Fe] -- [Fe/H] space around the four sub-components: \textit{Eos} (purple), Splash (turquoise), \textit{Aurora} (blue), GSE (red). The leftmost panel of Fig.~\ref{fig:fig9} shows these selections, the middle panel shows the energy distributions of the four selections, and the rightmost panel shows the ratio of the \textit{Eos} energy distribution to the other three energy distributions. From this figure we see that \textit{Eos} is most similar to the Splash. This similarity also suggests that \textit{Eos} is one continuous sequence of evolution up through the Splash stars, beyond [Al/Fe]~$\sim 0.2$ dex. This idea is reinforced by the appearance of component HE2, the small tail below [Al/Fe]~$\sim 0.2$ in component LE1, and the difference in the behaviour of the [Al/Fe]-rich ends of LE1 and LE2.

\begin{figure*}
\centering
\begin{subfigure}[b]{0.7\textwidth}
   \includegraphics[width=1\linewidth]{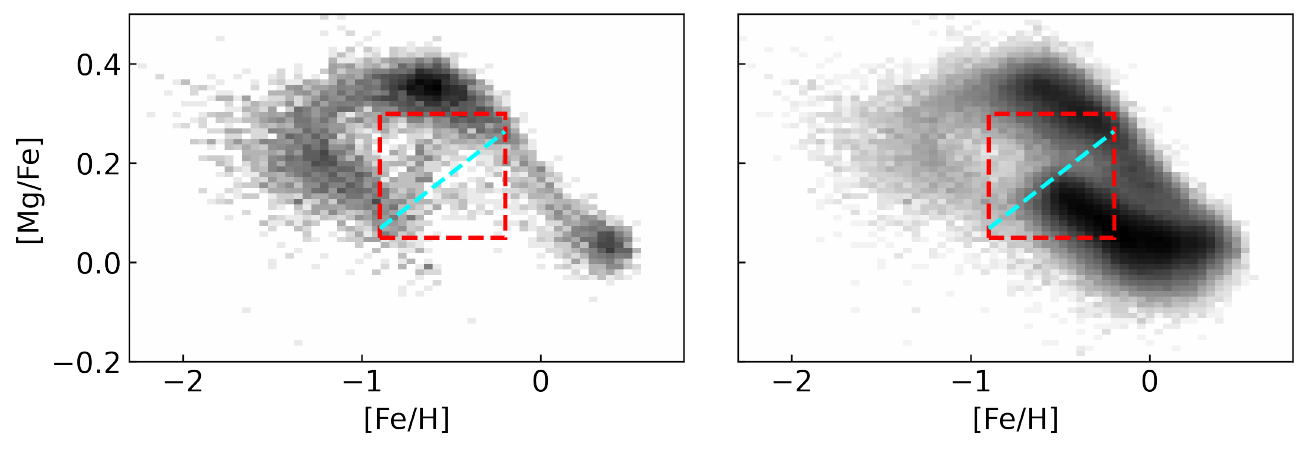}
   \label{fig:Ng1} 
\end{subfigure}

\begin{subfigure}[b]{0.8\textwidth}
   \includegraphics[width=1\linewidth]{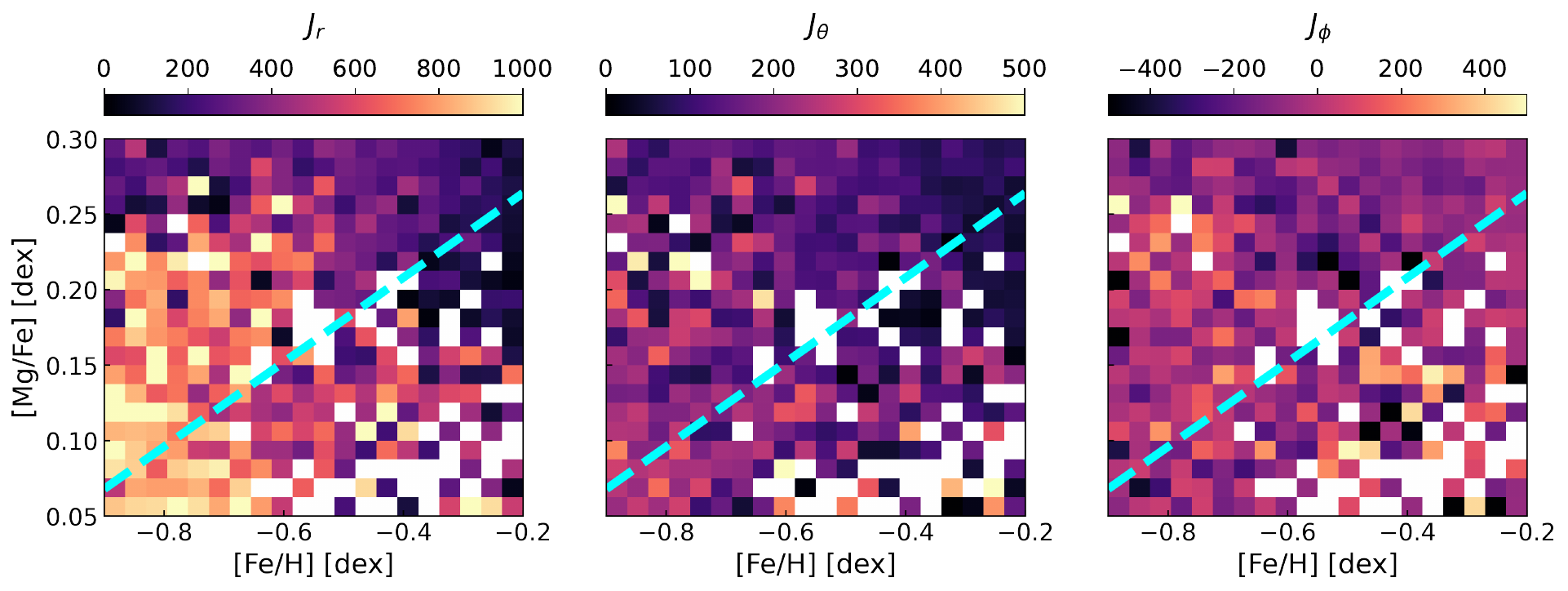}
\end{subfigure}

\caption[]{The region of the data occupied by the \textit{Eos} Mg bifurcation, coloured by different properties. The top panel highlights the bifurcation, i.e. two branches connecting the accreted population and the \textit{in-situ} population from [Fe/H]$\sim-0.9$ to [Fe/H]$\sim-0.2$. In the bottom 3 panels, we show binned mean values of various properties in this region. Specifically, we show (from left to right) radial action, angular action, and azimuthal action. The light blue dashed line roughly demarcates the boundary between the two branches. }
\label{fig:fig10}
\end{figure*}

When examined in [Mg/Fe] -- [Fe/H] space, \textit{Eos} exhibits a bifurcation. This was first noted by \citet[][]{myeong2022milky}, but is also evident in the bottom panel of HE2 in Fig.~\ref{fig:fig5}, as well as the bottom panel of Fig.~\ref{fig:fig2}. As of yet, there is no explanation for this. We investigate the bifurcation by plotting these stars, coloured by different kinematic properties. While we looked at a large range of parameters (including eccentricity and chemical abundances), the only difference exhibited between the two branches of the bifurcation was in radial action, $J_r$. In Fig.~\ref{fig:fig10}, we plot the bifurcation coloured just by actions. We include $J_{\phi}$ and $J_{\theta}$ for completion and for comparison. The top two panels of the figure illustrate which stars we have selected, as well as how the lower branch of the bifurcation overlaps entirely with the thin disk (low-$\alpha$ track). The dashed red square highlights the region in which the bifurcation is seen, and the dashed cyan line shows the boundary between the two branches of the bifurcation. The bottom panels of the figure present a clear difference between the two branches in $J_r$, but no difference in the other actions. To crudely quantify the difference between two branches, we calculate the median and mean values of $J_r$ above and below the dotted cyan line (within the chemical boundaries of the plot). Evidently, these values will have contamination from the GSE (bottom left corner) and the Splash (top right corner). Nevertheless, we find that the upper branch has a median $J_r = 484$ kpc km/s, and the lower branch as a median $J_r = 322$ kpc. Likewise, the upper branch has a mean $J_r = 582$ kpc km/s and the lower branch has a mean $J_r = 413$ kpc km/s. The standard deviation for the upper branch, at 2204 kpc km/s, is much higher than the lower branch, at 489 kpc km/s. The upper branch evidently displays more of a bias for radial orbits while the lower branch is kinematically more similar to the disk. 

We suggest that the lower branch of \textit{Eos} could be a result of bar-related resonance, like those described by \citet[][]{dillamore2023stellar, dillamore2024radial}. This mechanism has been shown to pull halo stars onto more disk-like orbits, and so could explain the low values of radial action, yet high energies. In [Mg/Fe] -- [Fe/H] space, this branch seems to connect the accreted halo with the thin disk, offering insight into mixing between these two populations as a possible result of bar related mechanisms. Clearly more work must be done to support our hypothesis.

\subsection{Aurora}

Running almost parallel with the \textit{Eos} branch, in [Al/Fe]--[Fe/H] space, is another feature which we identify as \textit{Aurora}, the ancient quasi-spheroidal ``heart'' of the MW formed before the MW disk formed \citep[][]{belokurov2022dawn, rix2022poor}. 

\begin{figure}
    \centering
    \includegraphics[width=0.9\columnwidth]{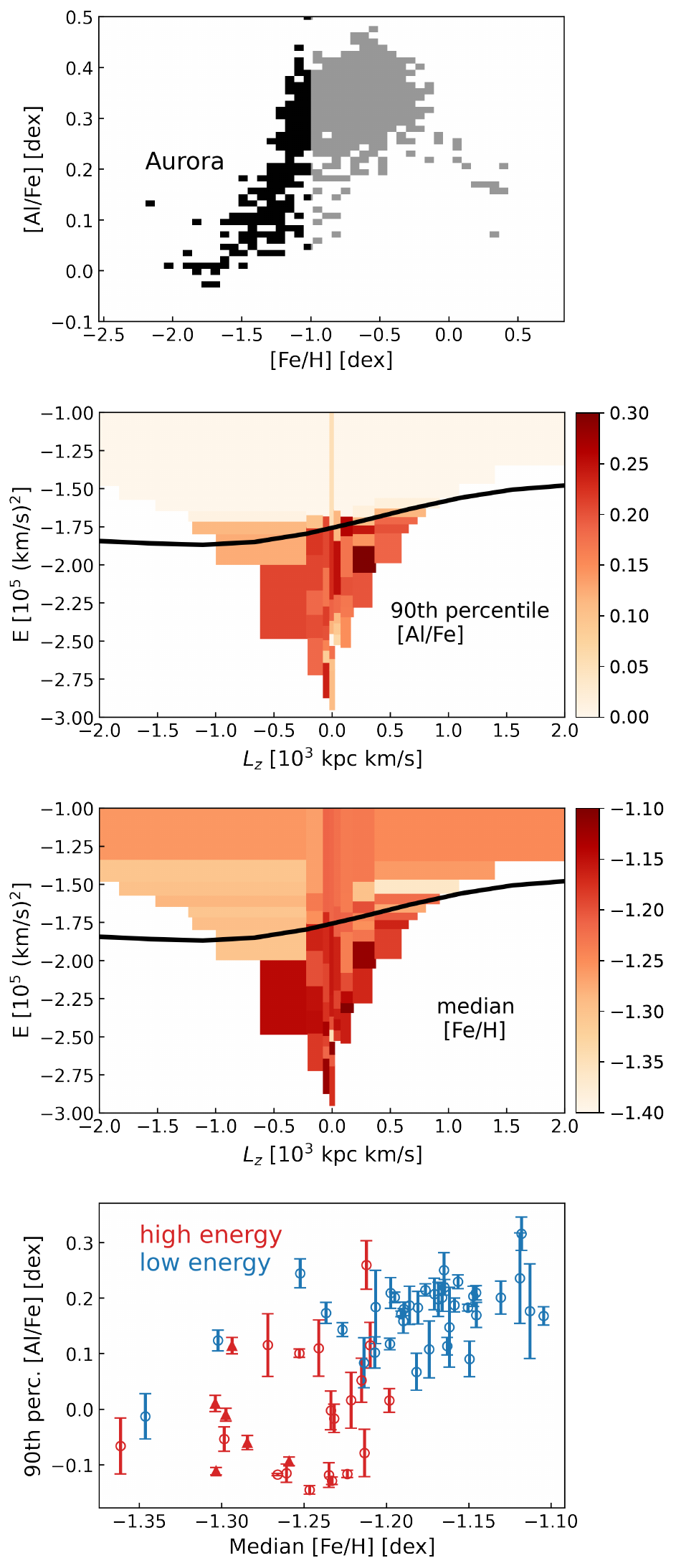}
    \caption{Chemical properties of the \textit{Aurora} branch mapped in $(E,L_z)$ space. \textit{Top:} The mask used for identifying bins that contain stars within the \textit{Aurora} branch (black), with full component shown for completeness (greyed out). \textit{Top-middle:} $(E,L_z)$ bins coloured by 90th percentile [Al/Fe] value, to indicate how high up the full branch each bin gets. Note the sharp cut-off at high energies. The black solid line is the pure chemical boundary. \textit{Bottom-middle:} $(E,L_z)$ bins coloured by median [Fe.H] value. The black solid line is the same. \textit{Bottom:} Median [Fe/H] plotted against 90th percentile [Al/Fe], with bootstrapped errorbars. We have coloured the points red if they are high energy (above the boundary) and blue if they low energy (below the boundary). Note the increasing trend of [Al/Fe] with [Fe/H]. The six retrograde $(E,L_z)$ bins that have distinctly low median [Fe/H] compared to other retrograde $(E,L_z)$ bins (as seen in the bottom-middle panel) are identified by filled triangles in the bottom panel.}
    \label{fig:fig11}
\end{figure}

To explore the chemical and kinematic properties of this population, we create a mask of the sequence using component LE1. The mask is defined as all bins in [Al/Fe] -- [Fe/H] space that are greater than $5\%$ of the maximum value of the component LE1, and are below [Fe/H] < -1.0 dex. The bins that satisfy this criteria are shown in black in the top panel of Fig.~\ref{fig:fig11}. We show the full component, but shade out bins not included at [Fe/H] > -1.0 dex. In each $(E,L_z)$ bin, we can then calculate the the 90th percentile value of [Al/Fe] within the mask, to estimate the length of the track between [Al/Fe]~$\sim 0.0$ and [Al/Fe]~$\sim 0.4$. We also calculate the median [Fe/H] value in each $(E,L_z)$ bin. We repeat these calculations for 1000 iterations, removing $\sqrt{N_{m'}}$ random stars from every bin in each iteration, to calculate means and standard deviations via bootstrap (where $N_{m'}$ is the number of stars in a given bin $m'$). In the lower three panels of Fig.~\ref{fig:fig11}, we plot these bootstrapped means and standard deviations. In the second panel we map the bootstrap mean 90th percentile of the [Al/Fe], in the third panel we map the bootstrap mean median [Fe/H] values, and in the fourth panel we plot these two against each other with errorbars shown. The errorbars presented in the fourth panel are shown to give an indication of whether or not the differences between the patches is on the order of the errorbars. 

The most striking result is the sharp cut-off of 90th percentile [Al/Fe] values at high energies in the second panel, as well as the patchwork nature of the [Al/Fe] values. The (mostly) small errorbars for these [Al/Fe] measurements indicates that we can trust these differences are genuine. This hints at a population of branches along the \textit{Aurora} branch all of different lengths, suggesting either (a) a collection of different star burst events that pile up on top of each other in [Al/Fe] -- [Fe/H] space, or (b) inhomogeneous mixing in the initial star forming gas. In Fig.~\ref{fig:fig5}, we see that this branch appears at a range of energies, seen in components HE2, LE1 and LE2. 

By inspecting the retrograde stars which straddle the \textit{in-situ}--accreted boundary in the middle two panels of Fig.~\ref{fig:fig11}, we notice that there is a patch of stars with [Al/Fe] and [Fe/H] values different from the surrounding regions. This suggests an evolution distinct from both the high energy and low energy low metallicity halo. We note that this region coincides with the kinematics where \textit{Thamnos}, a hypothesised accreted structure in the stellar halo \citep[e.g.][]{koppelman2019multiple, horta2021evidence, dodd2024character}, is expected to be found. Moreover, \textit{Thamnos} has been shown to appear along this [Al/Fe] branch which we have labelled as \textit{Aurora}. However, we also note that this patch of retrogade stars also kinematically and chemically aligns with $\omega$ Centauri \citep[e.g.][]{johnson2010omc,myeong2019evidence}. While we have attempted to clean out all GCs, it is possible that some extended tidal debris remains \citep[][]{ibata2019identification}, and is contaminating our picture. 

There also exists another, alternative explanation of the unusual chemical properties of the $E, L_z$ region where Thamnos allegedly resides. \citet{dillamore2024radial} demonstrate that interactions between the stellar halo and the Galactic bar produce pronounced sharp, narrow overdensities in the distributbsbssion of stars in the $E, L_z$ space. They reveal that these islands of resonantly trapped stars are not limited to the prograde portion of the angular momentum distribution. In the work of \citet{dillamore2024radial}, the resonance-induced ridges dominate the $L_z>0$ side, but can extend far into $L_z<0$ as well (cf also, Frakhoudi et al., in prep.). The most striking feature in their Figure 9 is the ridge corresponding to the 1:1 resonance (in $L_z>0$) which switches to 2:1 resonance for retrograde orbits. The location of this resonance appears to coincide, at least approximately, with the region claimed to contain the Thamnos sub-structure. In the simulations, the trapped $L_z<0$ stars contributing to this resonance had originally higher energies and higher angular momenta, having moved diagonally from higher up on the right towards the lower left (in the $E, L_z$ plane). As our analysis shows, the higher energy regions of the $E, L_z$ space are mostly composed of halo stars with lower [Al/Fe] values. Thus, in the proposed Thamnos region, the unusually low [Al/Fe] level could be due to the contribution of resonantly trapped stars from higher energies, rather than an actual accretion event.

\section{Summary \& Conclusions}\label{sec:results}

The aim of the so-called blind source separation (BSS) problem is to uncover a set of sources which make up a population of mixed signals, the most famous example of which is known as the cocktail party problem \citep[][]{cherry1953some}. In this work, we have treated the chemical mixtures within the Galactic stellar halo as a variant of (BSS), by applying non-negative matrix factorisation (NMF) to two selections of APOGEE DR17 stars. These selections include: (a) a selection based on kinematic cuts dubbed the ``chemically pure'' stellar halo, and (b) a selection based on chemical cuts dubbed the ``kinematically pure'' stellar halo. 

NMF is a machine learning technique, which was not previously used in the context of the stellar halo and which is particularly useful in uncovering distinct, overlapping, non-gaussian components in an unsupervised manner. At the same time, we found that the components would only be identified if they contributed over a broad range of the $E-L_z$ space and smaller features are often not identified. This is in line with the results of \citep[][]{chen2024galaxy} who found that only 2-4 major mergers can be reliably identified in the space of chemical abundances and integrals of motion. Therefore, we focus our attention on identifying major components of the Milky Way (MW), beginning with enforcing 2-components in our NMF model and then exploring 4-component decomposition.

Specifically, we use regions of $(E,L_z)$ space as our (chemically) mixed ``signal'' populations, in which we aim to find a selection of ``source'' components. In the 2-component model, for the chemically pure stellar halo, we found that the first component

\begin{itemize}
    \item occupied low energies ($E\lesssim -1.7\times10^5$ (km/s)$^2$),
    \item had median [Fe/H] = -0.55
    \item spanned -2 < [Fe/H] < 0.5, 
    \item dominated at [Al/Fe] > 0.0,
    \item spanned 0.0 < [Mg/Fe] < 0.4,
    \item contributed 66.5\% of the stellar halo,
    \item dominated the inner halo, at $(r,z,R) < (8.7, 3.0, 8.1)$ kpc,
\end{itemize}

broadly aligning with the \textit{in-situ} population of the MW. Likewise we found that the second component, broadly aligned with the accreted population of the MW, and

\begin{itemize}
    \item occupied high energies ($E\gtrsim -1.7\times10^5$ (km/s)$^2$),
    \item had median [Fe/H] = -1.15
    \item spanned -2 < [Fe/H] < -0.5, 
    \item dominated at [Al/Fe] < 0.0,
    \item spanned 0.0 < [Mg/Fe] < 0.4,
    \item contributed 33.5\% of the stellar halo,
    \item dominated the outer halo, at $(r,z,R) > (8.7, 3.0, 8.1)$ kpc.
\end{itemize}

For the 2-component, kinematically pure stellar halo, we found that the first component again occupies low energies, and has a wide distribution of eccentricities, from $e=0.2$ to $e=1.0$. While the second component again occupies high energies, and more strongly peaks at around an eccentricity of $e=0.8$. Based on the work by \citet[][]{amarante2022gastro}, this distribution could be used to infer the total mass of the GSE.

By plotting the fractional contribution of both sets of these components as a function of energy and distance, we find smooth, uninterrupted distributions that offer interesting insight into the make-up of the MW. Boundaries that delineate the accreted and \textit{in-situ} stellar halo (such as in \citep[][]{Belokurov2023nitrogen}) often suggest a hard energy separation between these two populations. Our work suggests, as one would naively expect, that a reasonable fraction ($\approx 10\%$) of the accreted debris can be found below this boundary and vice versa.

Using the components resulting from both the chemically pure dataset and the kinematically pure dataset, we define two new boundaries between the accreted and \textit{in-situ} stellar halo. We do this by finding the energy value (for a given $z$-angular momentum) at which the high energy components contribution falls below $50\%$. Using our boundary, we subsequently break each of the high energy (HE) and low energy (LE) components up into 4 further components, by applying two 2-component NMF models to these regions separately. In order of decreasing energy, we label these components HE1, HE2, LE1, LE2. In the 4-component model, we again explore the fractional contributions of each component as a function of distance and find that

\begin{itemize}
    \item component HE1 dominates the outer halo ($r > 9$ kpc) - chemically coinciding with the accreted debris (with median [Fe/H] = -1.22)
    \item component HE2 is found mostly at inner radii ($r < 9$ kpc), exhibiting chemical features overlapping with \textit{Eos} and Splash (with median [Fe/H] = -0.75),
    \item component LE1 is also found mostly at the inner radii ($r < 9$ kpc), exhbiting chemical features overlapping with \textit{Aurora} and Splash (with median [Fe/H] = -0.68),
    \item component LE2 dominates at inner radii ($r < 4$ kpc), and is either contamination from the bulge/bar, or some sign of a chemical gradient in the inner halo (with median [Fe/H] = -0.08).
\end{itemize}

These 4-components present a number of sub-structures which we examine further to understand their physical meaning, and the insight they provide into the MW's formation history. Mostly notably, within the [Al/Fe] -- [Fe/H] space of component HE2 and LE1, we find two branches that connect the accreted population with the Splash population. One branch sits just below [Fe/H] = -1.0 dex, and one branch sits just above. We identify the lower metallicity branch with the \textit{Aurora} population of {\it in-situ} pre-disk stars, and the higher metallicity branch with \textit{Eos}. Examination of these structures reveals a difference in $J_r$ between the two branches of the, already discovered, Mg bifurcation in \textit{Eos}. By isolating chemical bins that are only found within the Aurora branch, we map the 90th percentile [Al/Fe] value along the branch as a function of $(E, L_z)$ which reveals either (a) a range of overlapping evolutionary sequences along the Aurora branch, or (b) inhomogeneous mixing in the initial star-forming gas.

The two main halo components discussed here are derived through NMF analysis of APOGEE data in a completely unsupervised manner, with minimal assumptions. Despite this, they appear to align meaningfully with expectations for the in-situ and accreted contributions to the halo, as suggested by, for instance, the FIRE-2 numerical simulations. In all FIRE-2 Milky Way analogs, the in-situ fraction declines sharply beyond the Solar radius, consistent with our analysis. However, as Figure~\ref{fig:fire_sims} shows, the simulations reveal significant diversity in the fractional contributions of in-situ and accreted stars among Milky Way analogs. Many of the model galaxies exhibit shallower profiles for the in-situ contribution. Note however, the two simulated objects, \texttt{m12f} and \texttt{m12r}, which most closely match the Milky Way's age-metallicity relation within the FIRE-2 suite, display radial profiles strikingly similar to our measurements. 

One key advantage of NMF is that it allows for partial or complete overlap between sources. Indeed, the two halo components we recover exhibit noticeable overlap, especially at low metallicities (see Figure~\ref{fig:fig3}). However, despite this, the centroids of the components remain well-separated across the entire metallicity range studied, i.e., $-2<$[Fe/H]$<0.5$. Even at [Fe/H]$<-1.5$, the low-energy component forms a narrow, distinct track in both [Al/Fe]-[Fe/H] and [Mg/Fe]-[Fe/H] spaces. Previous studies, such as \citet{belokurov2022dawn}, used simple cuts to distinguish in-situ from accreted populations, but these became unreliable around [Fe/H]$\approx-1.5$, where the in-situ sequence drops below [Al/Fe]$=0$. Our analysis provides a more organic definition of the Aurora stellar component, the pre-disk in-situ population of the Milky Way, avoiding rigid abundance-based cuts. This reveals Aurora extending continuously to [Fe/H]$\approx-2$ and [Al/Fe]$\approx-0.3$, a region typically associated with the accreted population. This seamless transition of the ancient Milk Way sequence across the [Al/Fe]$=0$ boundary makes sense and is supported by chemical evolution models \citep[see e.g. Figure 2 in][]{horta2021evidence}.

The results of this work illustrate that NMF, which makes no assumptions about the distributions of the components it looks for
(other than non-negativity), is a powerful tool for uncovering complex structures in the chemo-dynamical space occupied by the Galactic stars. We hope that this work further encourages the use of BSS (and NMF) for studying the Galaxy.

\section*{Acknowledgements}

The authors thank the reviewer, Sten Hasselquist, for helpful comments that increased the clarity of our results. The authors would additionally like to thank the rest of the Cambridge streams group for useful conversations. EYD thanks the Science and Technology Facilities Council (STFC) for a PhD studentship (UKRI grant number 2605433). SGK also thanks Elana Kane for her expertise on blind sources.

We used FIRE-2 simulation public data  \citep[][]{Wetzel2023}, which are part of the Feedback In Realistic Environments (FIRE) project, generated using the Gizmo code \citep{hopkins15} and the FIRE-2 physics model \citep{hopkins_etal18}. 

\section*{Data Availability}

We make use of publicly available APOGEE DR17 data. The FIRE-2 simulation data used in this study is available at 
\url{http://flathub.flatironinstitute.org/fire}.



\bibliographystyle{mnras}
\bibliography{bss} 




\appendix

\section{Inner-galaxy contamination}\label{sec:appendix1}

\begin{figure}
    \centering
    \includegraphics[width=0.98\columnwidth]{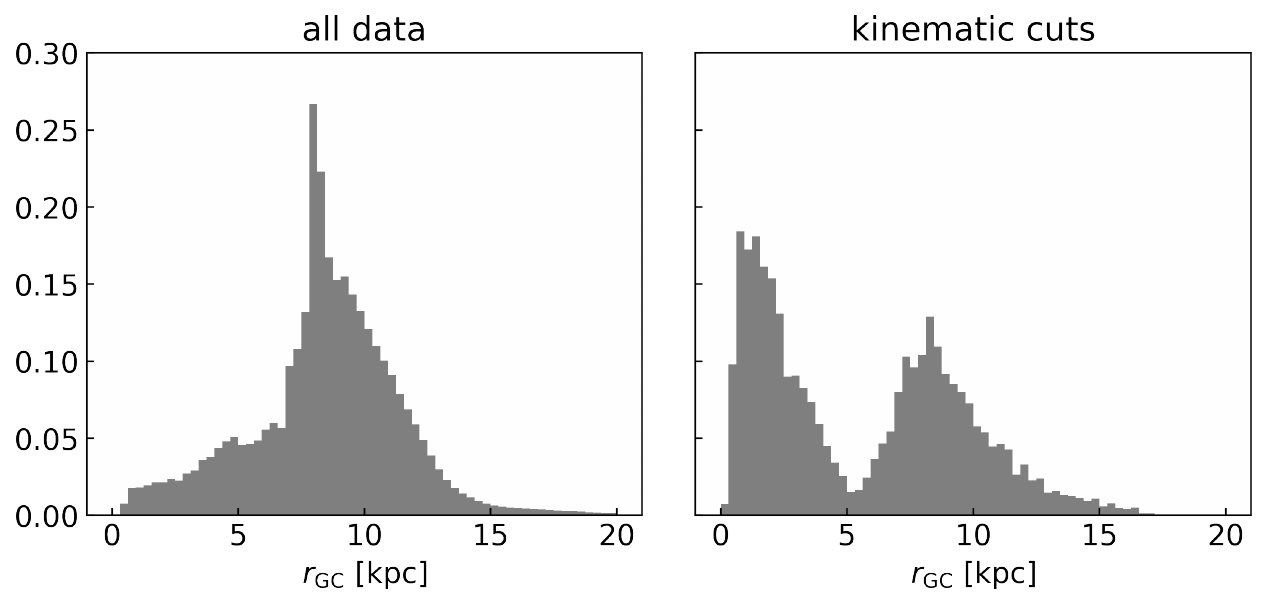}
    \caption{Galactocentric radial distribution of stars from the entire sample (left), and the ``chemically pure'' sample (right) which is obtained by kinematic cuts. It is evident from this figure that many of the stars in the chemically pure sample are confined to the inner galaxy ($r_{\rm gc} < 5$ kpc). This distribution motivates appendix 1, where we examine the chemically pure sample with these inner stars removed.}
    \label{fig:appendix1}
\end{figure}

\begin{figure}
    \centering
    \includegraphics[width=0.98\columnwidth]{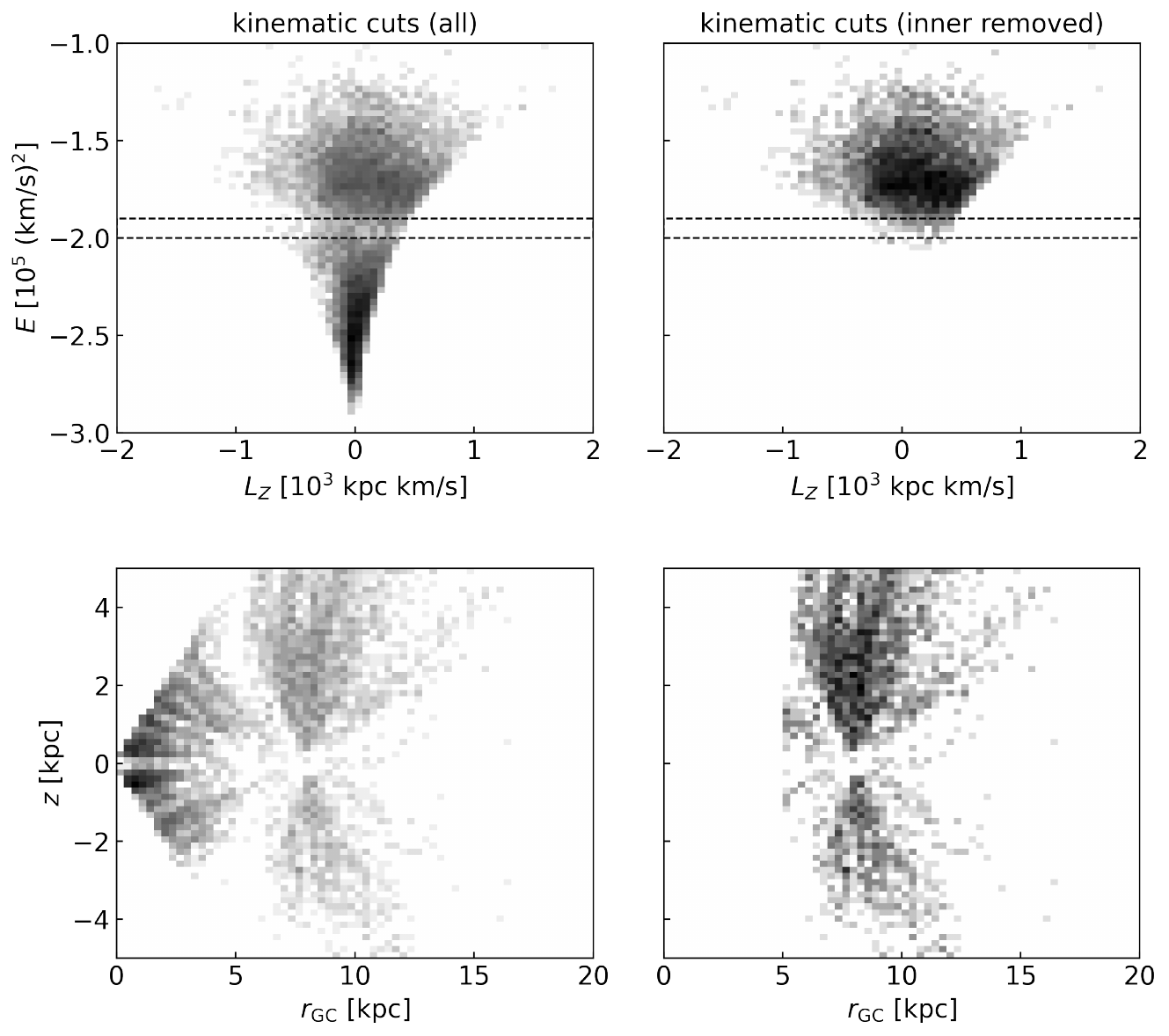}
    \caption{Spatial and integral of motion distribution comparison of the chemically pure sample before and after removing the inner 5 kpc stars. The horizontal dashed line guides the readers eye to the APOGEE selection effect feature.}
    \label{fig:appendix1a}
\end{figure}

\begin{figure}
    \centering
    \includegraphics[width=0.95\columnwidth]{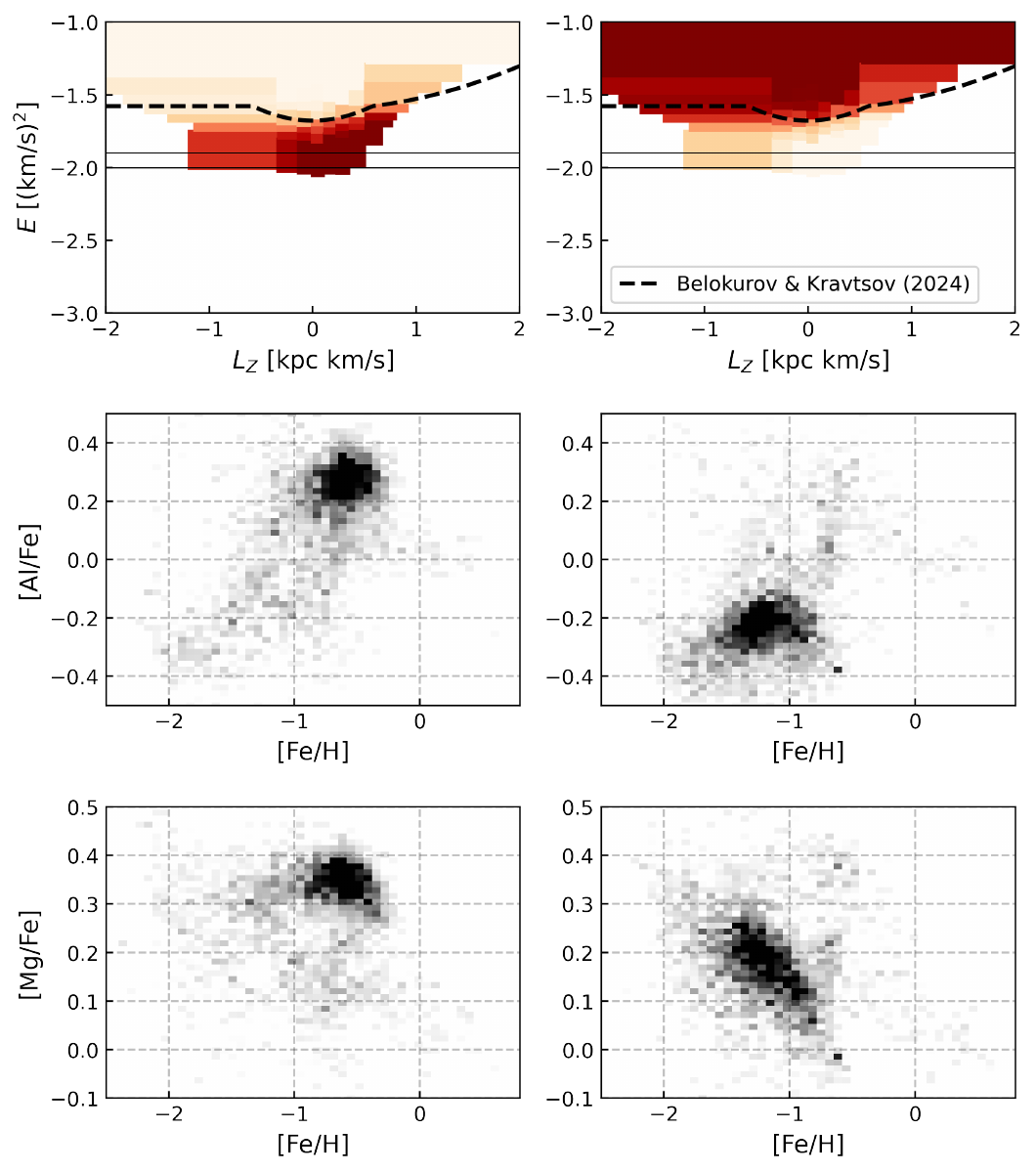}
    \caption{The resulting components from a 2-component NMF applied to the chemically pure stellar halo data, with the stars found within 5 kpc of the galactic centre removed. The tow row presents energy and z-angular momentum space, the middle row presents aluminium chemical information, whereas the bottom row presents magnesium chemical information.The horizontal dashed line in the top row guides the readers eye to the APOGEE selection effect feature, and how it is not aligned with the NMF derived boundary.}
    \label{fig:appendix2}
\end{figure}

\begin{figure}
    \centering
    \includegraphics[width=0.95\columnwidth]{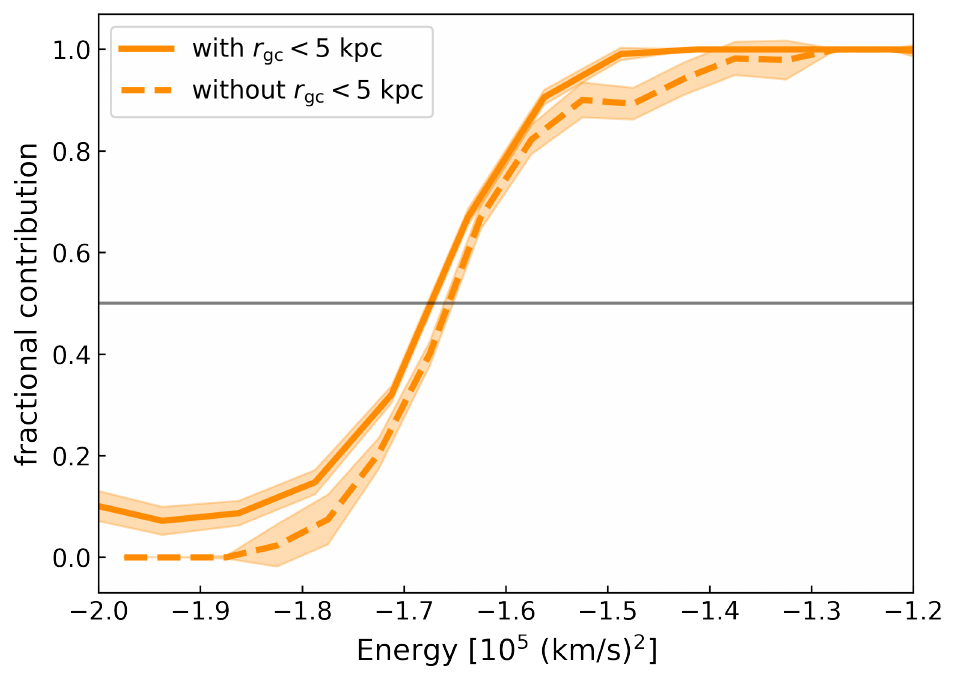}
    \caption{Comparison of the fractional contributions of the accreted components from two 2-component NMF models. One takes the original chemically pure sample as input, and the other takes the chemically pure sample with the inner 5 kpc removed. The orange solid line shows the original sample, while the orange dashed line shows the sample with the inner galaxy stars removed. From this figure it is evident that there is very little difference in the transition point to accreted domination.}
    \label{fig:appendix3}
\end{figure}

\begin{figure*}
    \centering
    \includegraphics[width=0.99\textwidth]{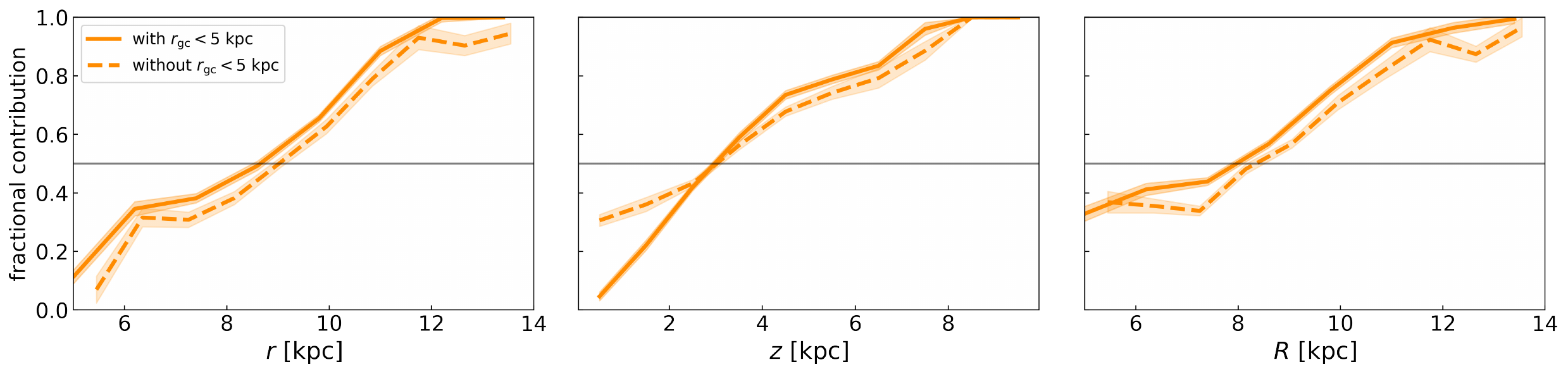}
    \caption{Comparison of the fractional contribution of the accreted components from two 2-component NMF models as a function of galactocentric spherical radius ($r$), galactocentric height ($z$) and galactocentric cylindrical radius ($R$). The orange solid line shows the original sample, while the orange dashed line shows the sample with the inner galaxy stars removed.}
    \label{fig:appendix4}
\end{figure*}

In this section, we present the results of excluding the inner stars of our so-called chemically pure sample (i.e. those with $r_{\rm GC} < 5$ kpc). Fig.~\ref{fig:appendix1} shows the distribution of stars in the chemically pure sample as a function of galactocentric radius, and highlights the bimodal nature of the distribution of stars. Specifically, of the 13147 total stars in the kinematic sample, 6529 of the them are within 5 kpc of the galactic centre, and 6618 of them are not. Prior to our kinematic selection cuts, and after quality cuts to the APOGEE data, 37572 stars were found within 5 kpc of the galactic centre and 250176 were found beyond 5 kpc. This means $13\%$ of the original APOGEE stars were found in the inner galaxy, while almost $50\%$ of ``chemically pure'' sample stars are in the inner galaxy. Fig.~\ref{fig:appendix1a} shows the integral of motion space and $(r,z)$ space the chemically pure sample, before and after removing the inner 5 kpc.

To investigate the impact of removing the inner 5 kpc from the chemically pure sample, we re-run the exact same experiments described in Sec.~\ref{sec:two_component_model}, with these inner stars removed. The components resulting from a 2-component NMF applied to the sample with the inner galaxy removed are presented in Fig.~\ref{fig:appendix2}. We can see that the boundary's location is quite unchanged from the original chemically pure sample. Moreover, in both Fig.~\ref{fig:appendix1a} and Fig.~\ref{fig:appendix2} we highlight the location of the APOGEE selection function artifact to illustrate how it is different from the NMF derived boundary.

Similarly, we repeat the experiments described in Sec.~\ref{sec:contribution_as_a_function_of_energy} and Sec.~\ref{sec:contribution_as_a_function_of_distance} to find the contribution of the two components as a function of energy and as a function of distance. This allows us to compare the points at which the halo becomes accretion dominates as a function of these two metrics, and compare with the original chemically pure sample. The results of these experiments is presented in Fig.~\ref{fig:appendix3} and Fig.~\ref{fig:appendix4}. We can see that there is a very minimal difference between the transition points both in spatial distance and in energy. Namely, the transition point shifts from $(r,z,R) \approx (8.7, 3.0, 8.1)$ kpc to $(r,z,R) \approx (9.0, 3.0, 8.2)$ kpc.

Despite the similarity of the transition point (both energy and spatial) after the removal the inner galaxy stars, the percentage contribution of the accreted and the \textit{in-situ} population to the total halo stars can interpreted to be very different. Specifically, the low-energy component (approximately the \textit{in-situ} population) contributes only $47.06\%$, whereas the high-energy component (approximately the accreted population) contributes the remaining $52.94\%$.

\section{Spatial distribution of \textit{Eos}}

In this section we present some additional information regarding the behaviour of the two branches of \textit{Eos} that are most visible in [Mg/Fe] -- [Fe/H] space. To clarify the spatial distribution of these two populations, we recreate Fig.~\ref{fig:fig10} with the chemical space coloured by galactocentric height and radius. Visually, it appears that there is no dramatic difference between the height and radius distributions. To quantify the difference, we calculate the mean radius and mean height in the two regions below the cyan line (not including the contamination from the high-$\alpha$ sequence at [Mg/Fe] > 0.2 dex). While the mean height was 1.4 kpc for both regions, the upper branch had a mean radius of 6.8 kpc and the lower branch had a mean radius of 8.3 kpc. We also examined the galactic latitude, galactic longitude, [Al/Fe] and eccentricity distributions of the two branches, and found negligible difference.

\begin{figure}
    \centering
    \includegraphics[width=0.99\columnwidth]{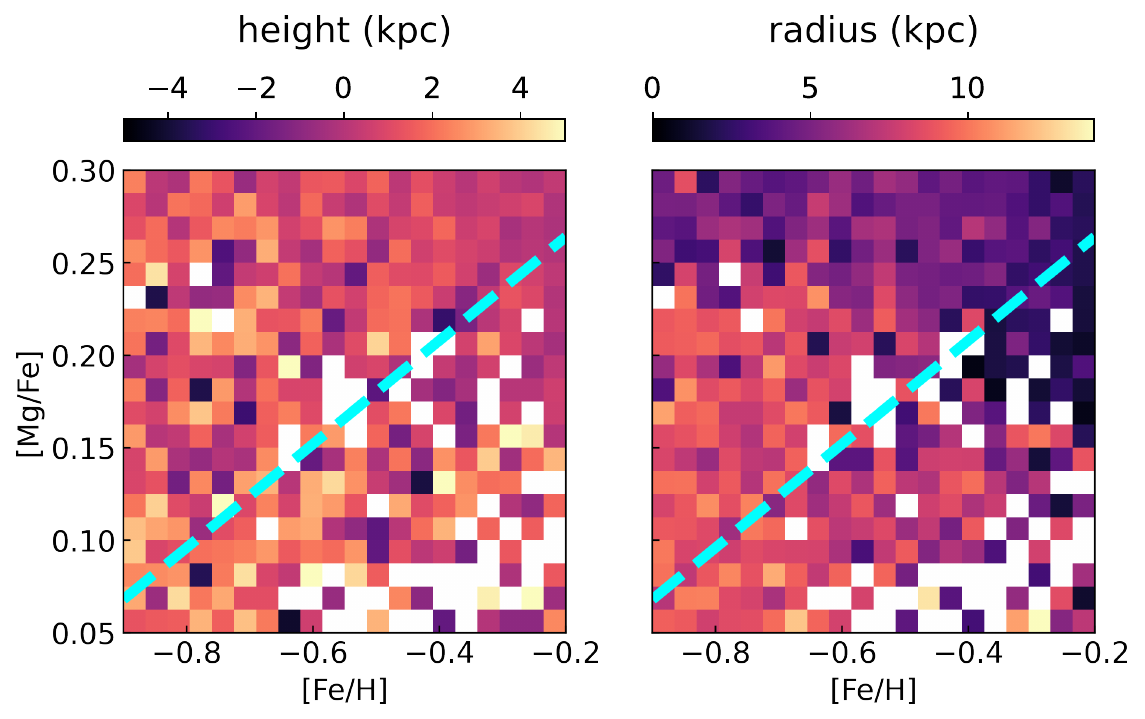}
    \caption{Re-examination of the chemical space in Fig.~\ref{fig:fig10}, but coloured by galactocentric height and radius. There is little visible difference between the two branches. The mean height was 1.4 kpc for both the upper and lower branches, while the upper branch had a mean radius of 6.8 kpc and the lower branch had a mean radius of 8.3 kpc.}
    \label{fig:appendix6}
\end{figure}


\bsp	
\label{lastpage}
\end{document}